%% file: v2.tex
\documentclass[11pt]{article}
\pdfoutput=1
\usepackage[utf8x]{inputenc}
\usepackage{amssymb,amsmath,mathrsfs,enumerate,appendix}
\usepackage{graphicx,rotate,multicol, multirow}
\usepackage{graphicx}

\graphicspath{plots/}
\usepackage{cite}
\usepackage{braket}
\usepackage[normalem]{ulem}
\usepackage{wrapfig}
\usepackage[svgnames,table]{xcolor}
\usepackage[textfont=it,labelfont=bf]{caption}
\usepackage{bm}
\usepackage[colorlinks=true,
linkcolor=red,
urlcolor=blue,
citecolor=blue]{hyperref}
\usepackage{lineno}
\usepackage{color}
\usepackage{booktabs}
\usepackage{makecell}
\usepackage{subcaption}
\usepackage{bbold}
\usepackage{bm}
\usepackage{verbatim}
\long\def\rpl#1!!#2!!{\textcolor{red}{#1} \textcolor{blue}{#2}}
\def\@seccntformat#1{\@ifundefined{#1@cntformat}%
	{\csname the#1\endcsname\quad}%      default
	{\csname #1@cntformat\endcsname}%    enable individual control
}

\def\bar{\overline}
\def\tilde{\widetilde}

\makeatother

\newcommand{\bea}{\begin{eqnarray}}
\newcommand{\eea}{\end{eqnarray}}

\newcommand{\be}{\begin{equation}}
\newcommand{\ee}{\end{equation}}
\def\beq{\begin{equation}}
\def\eeq{\end{equation}}
\newcommand{\ba}{\begin{eqnarray}}
\newcommand{\ea}{\end{eqnarray}}
\def\ifmath#1{\relax\ifmmode #1\else $#1$\fi}

\def\bu {{\bm{u}}}
\def\bd {{\bm{d}}}
\def\bl {{\bm{\ell}}}

\evensidemargin=\oddsidemargin
\def\Eqn#1{Eq.\ (\ref{#1})}
\def\Eqs#1#2{Eqs.\ (\ref{#1}) and (\ref{#2})}

\allowdisplaybreaks
\title{\Large\bf 
	Probing an up-phobic Higgs at colliders: the role of $b$-associated production in the democratic 3HDM
}

\author{
	\sf 
	Dipankar Das$^{a,}$\footnote{d.das@iiti.ac.in},
	Teesha Khandelwal$^{b,}$\footnote{ph24d009@smail.iitm.ac.in}, Ipsita Saha$^{b,}$\footnote{ipsita@iitm.ac.in}, 
	\\[3mm]
    \small\em
	$^a$Department of Physics, Indian Institute of Technology Indore, Khandwa Road, Simrol,
	Indore 453 552, India \\
	\small\em
	$^b$Department of Physics, Indian Institute of Technology Madras, Chennai 600036, India \\
}

\date{}

\begin{document}
	
%\preprint{CFTP/22-004}  
%\pagewiselinenumbers
	
\maketitle
\renewcommand*{\thefootnote}{\arabic{footnote}}
\setcounter{footnote}{0}
	
\begin{abstract}
We show that there exists a theoretically well-motivated Higgs sector in which the conventional heavy-Higgs production mechanisms are suppressed. 
In this limit, the $b$-associated production channel assumes the role of the primary production mechanism. We demonstrate that the democratic three Higgs-doublet model (3HDM) provides a viable beyond the SM (BSM) framework in which this limit can be realized. We further show that existing $b$-associated searches provide the most sensitive LHC probe of this 3HDM scenario in the up-phobic limit and derive the corresponding constraints on its parameter space. We also present projections for future collider searches, demonstrating their potential to probe further regions of the model's up-phobic parameter space.
% in contrast to the minimal 2HDM, where such a limit can not be attained. 

\end{abstract}
	
	\maketitle
	
\section{Introduction}
\label{sec:intro}
%%%%%%%%%%%%%%%%%%%%%%%%%%%%%%%%%%%%%%%%%%%%%%%%%%%%%%%%%%%%%%%%%%%%%
%%%%%%%%%%%%%%%%%%%%%%%%%%%  Introduction  %%%%%%%%%%%%%%%%%%%%%%%
%%%%%%%%%%%%%%%%%%%%%%%%%%%%%%%%%%%%%%%%%%%%%%%%%%%%%%%%%%%%%%%%%%%
%The discovery of a scalar particle with a mass around 125 GeV at the Large Hadron Collider (LHC) has completed the particle spectrum predicted by the Standard Model (SM) 
%while simultaneously opening a new avenue for exploring the dynamics of electroweak symmetry breaking. Measurements of the couplings of this Higgs boson have so far shown 
%remarkable agreement with the corresponding SM predictions, indicating that the observed scalar behaves, to a very good approximation, like the Higgs boson anticipated 
%within the SM. 
%Nevertheless, the scalar sector remains one of the least constrained sectors of the theory, %The existence of additional scalar degrees of freedom is neither 
%theoretically excluded nor experimentally ruled out, 
%making 

Extensions of the Standard Model (SM) Higgs sector are one of the most actively explored directions in physics beyond the Standard Model (BSM) which allude to the intriguing 
possibility that the 125 GeV scalar found at the LHC might only be the first one in a series of many others to follow.
Among the simplest and most extensively studied possibilities are models containing multiple Higgs doublets for they keep the tree-level value of the electroweak (EW) $\rho$-parameter unaltered. 
%Such models arise naturally in several theoretical frameworks 
%addressing questions ranging from the origin of flavor and CP violation to supersymmetric theories and models with spontaneous symmetry breaking at higher scales. 
%In addition to predicting an enriched scalar spectrum, multi-Higgs-doublet models provide a versatile framework in which different fermion sectors may couple to different 
%scalar multiplets, thereby giving rise to novel Yukawa structures and distinctive collider signatures. These features make them particularly attractive from the perspective 
%of LHC phenomenology.
%
While two Higgs-doublet models (2HDMs)~\cite{Branco:2011iw,Bhattacharyya:2015nca} have been investigated in considerable detail, models with three Higgs-doublets offer a significantly richer phenomenology which are yet to be explored. 
In particular, the democratic~\cite{Cree:2011uy,Das:2019yad,Chakraborti:2021bpy} or Type-$Z$ three Higgs-doublet model (3HDM)~\cite{Boto:2021qgu} possesses a unique Yukawa structure in which each fermion sector receives its mass exclusively 
from one Higgs doublet. Specifically, one scalar doublet couples only to up-type quarks, another only to down-type quarks, while the third couples exclusively to charged 
leptons\footnote{This differs from the class of models known as the “Private Higgs” models~\cite{Porto:2007ed,Porto:2008hb}, in which each individual massive fermion obtains its mass from a distinct Higgs doublet, resulting in a total of nine Higgs doublets for nine fermions.}. Such a flavor assignment naturally avoids tree-level scalar mediated flavor-changing neutral currents (FCNCs) through the principle of natural flavor conservation (NFC)~\cite{Glashow:1976nt}.
%while simultaneously 
%enlarging the space of possible Yukawa interactions compared with conventional 2HDMs.

An intriguing consequence of this Yukawa structure is the existence of special regions of parameter space where one family of fermionic Yukawa couplings to a nonstandard neutral 
scalar can vanish identically. Depending on the values of the scalar mixing angles, one may realize an up-phobic limit, where the nonstandard neutral scalar has no couplings to up-type 
quarks, including the top quark; a down-phobic limit, where couplings to down-type quarks disappear; or a leptophobic limit, where couplings to charged leptons are absent. 
These limits are unique to the democratic Yukawa structure and do not possess direct analogues in simpler Higgs-sector extensions such as 2HDMs. They therefore provide an interesting 
laboratory for studying collider signatures that differ substantially from the conventional expectations for extended Higgs sectors.

In this work we concentrate on the up-phobic limit. The absence of couplings to up-type quarks has particularly important phenomenological consequences because it strongly suppresses the dominant production mechanism for neutral heavy Higgs bosons at hadron colliders, namely gluon fusion mediated by top-quark loops. Consequently, the production rates 
and decay patterns of the nonstandard scalar become qualitatively different from those usually considered in heavy-Higgs searches. Since many existing LHC analyses have 
been designed assuming substantial gluon-fusion production, it becomes important to reassess the sensitivity of present searches in scenarios where this production channel is substantially diminished.
%\sout{either strongly suppressed or completely absent}

A further simplification, and one that is well motivated by current Higgs measurements, is provided by the 3HDM alignment limit~\cite{Das:2019yad}. The observed properties of the 125 GeV Higgs 
boson indicate that its couplings to gauge bosons and fermions are very close to their SM values, suggesting that the scalar sector lies near this alignment regime. In 
this limit, the lightest CP-even Higgs possesses SM-like interactions, while the heavier CP-even neutral scalars lose their tree-level couplings to electroweak gauge boson 
pairs, namely $WW$ and $ZZ$. Consequently, when the alignment limit is combined with the up-phobic Yukawa configuration, the heavy CP-even scalar becomes simultaneously decoupled from the top quark and from the massive electroweak gauge bosons. This curbs not only gluon-fusion production but also eliminates vector-boson-fusion and Higgs-strahlung or vector-boson associated production 
channels, thereby suppressing the production mechanisms that ordinarily dominate neutral Higgs searches at the LHC.

This immediately raises a natural phenomenological question: through which processes can such a scalar still be produced efficiently at the LHC? The answer lies in its couplings to down-type quarks and charged leptons, which remain unsuppressed in the up-phobic limit. In particular, production in association with bottom quarks, followed by the decay of the heavy neutral scalar into a pair of tau leptons, emerges as one of the most promising discovery channels. Unlike in more generic Higgs-sector extensions, where $b$-associated production often complements gluon-fusion searches, in the present scenario it effectively becomes the primary, and in an important region of parameter space the exclusive, probe of the nonstandard neutral scalar $(S)$. This observation substantially enhances the phenomenological significance of dedicated $(b\bar bS)$ searches performed by the LHC experiments.

Motivated by this observation, we investigate the capability of existing $b$-associated heavy-Higgs searches to constrain the parameter space of the democratic 3HDM in 
the up-phobic regime. After briefly reviewing the model, we identify the conditions under which the up-phobic limit is realized and 
discuss its interplay with the alignment limit. We then reinterpret the current LHC searches for $(pp\to b\bar b S)$, followed by $(S \to\tau^+\tau^-)$~\cite{ATLAS:2020zms}, to quantify the resulting 
bounds on the mass and couplings of the nonstandard neutral scalar. 
%Our analysis demonstrates that, once the conventional production channels are absent, the 
%$(b\bar bH/A(\tau^+\tau^-))$ channel assumes a uniquely important role in testing this otherwise elusive region of the parameter space. We therefore argue that the ongoing and 
%future improvements in dedicated (b)-associated Higgs searches will provide one of the most direct experimental windows into the up-phobic limit of the democratic 
%three-Higgs-doublet Model.
%

\section{The Model in brief}
\label{sec:model}
The Type-$Z$ or democratic Yukawa structure is characterized by the scenario when each type of charged fermions receives their masses from 
its own dedicated scalar doublet, conveniently denoted as $\phi_\bu$, $\phi_\bd$ and $\phi_\bl$ respectively~\footnote{In earlier works on Type-$Z$ or democratic 3HDM,
$\phi_\bu$, $\phi_\bd$ and $\phi_\bl$ were denoted as $\phi_3$, $\phi_2$ and $\phi_1$ respectively~\cite{Das:2019yad,Boto:2021qgu}. Our current notational alteration is intentional as it directly indicates the type of fermion that a particular doublet gives mass to.}. 
Such an arrangement of Yukawa couplings is usually ensured by imposing an extra $Z_3$ or $Z_2 \times Z_2^\prime$ symmetry in addition to the 
$SU(2)_L \times U(1)_Y$ gauge symmetry. For the version with $Z_3$ symmetry, the different fermion and scalar fields are assumed to be
transforming as follows~\cite{Das:2019yad}:
 \begin{subequations}
 	\begin{eqnarray}
 		\phi_\bl \to \omega \phi_\bl, & \bl_R^0 \to \omega^2 \bl_R^0, \\
 		% \nonumber\\
 		\phi_\bd \to \omega^2 \phi_\bd, & \bd_R^0 \to \omega \bd_R^0, 
 %		\phi_u \to \phi_u, &\ u_R \to u_R.
 	\end{eqnarray}
 \label{e:z3symm}
 \end{subequations}
where $\omega = e^{2\pi i/3}$, and $\bd_R^0$ and $\bl_R^0$ denote the right-handed down-type quark and charged lepton, respectively. 
The other fields, including the third scalar doublet $\phi_\bu$ do not transform under the $Z_3$ symmetry. 
The superscript $0$ emphasizes that the fermionic fields are in the gauge basis, which should be distinguished from the fields in the 
physical mass basis.
As an alternative, one may also choose to impose a $Z_2 \times Z_2^\prime$ symmetry acting as follows~\cite{Ivanov:2011ae,Keus:2013hya,Ivanov:2014doa,Das:2022gbm,Boto:2023nyi}:
 \begin{subequations}
	\begin{eqnarray}
	{\rm under}~Z_2 : &	\phi_\bl \to - \phi_\bl, & \bl_R^0 \to - \bl_R^0, \\
		% \nonumber\\
	{\rm under}~Z_2^\prime : &	\phi_\bd \to - \phi_\bd, &\ \bd_R^0 \to - \bd_R^0. 
		%		\phi_u \to \phi_u, &\ u_R \to u_R.
	\end{eqnarray}
\label{e:z2symm}
\end{subequations}
As before, the other fields do not transform under the $Z_2 \times Z_2^\prime$ symmetry.
Both the symmetries in \Eqs{e:z3symm}{e:z2symm} will lead to the following Yukawa Lagrangian:
\begin{eqnarray}
	{\mathscr L}_Y = - \sum_{i,j =1}^{3}\left[\bar{Q}_{L,i}^0 Y^\bd_{ij} \phi_\bd d_{R,j}^0 + \bar{Q}_{L,i}^0 Y^\bu_{ij} \tilde{\phi}_\bu u_{R,j}^0  + \bar{L}_{L,i}^0 Y^\bl_{ij} {\phi}_\bl \ell_{R,j}^0  \right] + {\rm h.c.},
\end{eqnarray}
%where $Q_{L}^0$ and $L_{L}^0$ denote the left-handed quark and lepton doublets respectively, and $\tilde{\phi}_{\bu} = i\sigma_{2} \phi_{\bu}^*$ with $\sigma_{2}$ being the 2nd Pauli matrix.\pink{ $Y^{\bu,\bd,\bl}$ are corresponding Yukawa matrices.}
where $Q_{L}^0$ ($L_{L}^0$) and $u_R^0$ denote the left-handed quark (lepton) doublets and right-handed up-type quarks respectively, and $\tilde{\phi}_{\bu} = i\sigma_{2} \phi_{\bu}^*$ with $\sigma_{2}$ being the 2nd Pauli matrix. $Y^{\bu,\bd,\bl}$ are corresponding Yukawa matrices.
Note that, the phenomenological discussions presented in this article depend primarily on the above Yukawa Lagrangian and
not on the exact detail of the symmetry from which it follows. In fact, the difference between the $Z_3$ and $Z_2 \times Z_2^\prime$
 versions of the Type-$Z$ 3HDM will arise solely from the scalar potential~\cite{Das:2022gbm,Boto:2023nyi} which will not be very relevant for our current
 phenomenological analysis.
 
 Now, we will establish the notations for the physical Higgs bosons which will emerge from the scalar sector.
 For a detailed discussion of the scalar sector of a democratic 3HDM, we refer the reader to appendix~\ref{sec:appendix},
 	where we present the detailed analysis of a 3HDM scalar potential with a softly-broken $Z_2 \times Z_2^\prime$ symmetry.
 %we refer the reader to Refs.~\cite{Das:2019yad,Chakraborti:2021bpy}.
 Here, we will be brief and discuss only those parts relevant for understanding the Yukawa couplings introduced later,
 in a self-contained manner. 
 
 After the electroweak symmetry breaking, the scalar doublets are decomposed in terms of their component fields as
 \begin{eqnarray}
 	\phi_f & = & 	\frac{1}{\sqrt{2}}\left(\begin{array}{c}
 	\sqrt{2}	\omega_f^+ \\
 	 v_f + h_f + i z_f
 	\end{array}\right), \label{e:expand_fields}
 \end{eqnarray}
where $f \equiv \bu,\bd,\bl$ and $v_f$ denotes the vacuum expectation value (VEV) of the corresponding scalar doublet.
The VEVs are further parametrized as~\footnote{Throughout this paper, the subscripts in boldface denote the type of fermions whereas the subscripts in the normal font refer to a particular fermion. For example,
$\kappa_\bu$, as introduced later, will denote the common coupling modifier for all up-type quarks, whereas $\kappa_u$ denotes the same for up quark alone. }
\begin{eqnarray}
	v_\bu = v \sin \beta_2; \quad v_\bd = v \sin \beta_1 \cos \beta_2; \quad v_\bl = v \cos \beta_1 \cos \beta_2\,, \label{e:vev-parametrization}
\end{eqnarray}
where $v^2 = v_\bu^2 + v_\bd^2 + v_\bl^2$ denotes the square of the total electroweak VEV. To obtain the physical scalars, it is useful to define the following
$3\times3$ orthogonal matrix,
\begin{eqnarray}\label{e:obeta}
	{\cal O}_{\beta} =
	\begin{pmatrix} \cos\beta_2 \cos\beta_1 & \cos\beta_2 \sin\beta_1
		&  \sin\beta_2 \\
		-\sin\beta_1 & \cos\beta_1  &  0  \\
		-\cos\beta_1 \sin\beta_2 & -\sin\beta_1\sin\beta_2 & \cos\beta_2 
	\end{pmatrix}.
\end{eqnarray}
The charged and the neutral Goldstone bosons, denoted by $G^\pm$ and $G^0$ respectively are then 
obtained via the following rotation:
\begin{equation}
	\begin{pmatrix} G^\pm\\ H_1^{\prime \pm} \\ H_2^{\prime \pm}\end{pmatrix} =
	\mathcal{O}_\beta
	\begin{pmatrix} w_\bl^\pm\\ w_\bd^\pm \\ w_\bu^\pm\end{pmatrix}\; , \quad
	\quad
	\begin{pmatrix} G^0\\ A_1^\prime \\ A_2^\prime \end{pmatrix} =
	\mathcal{O}_\beta 
	\begin{pmatrix} z_\bl \\ z_\bd \\ z_\bu \end{pmatrix} .
	\label{e:betarot1-cpodd}
\end{equation}
The physical charged scalars $(H_{1,2}^\pm)$ and pseudoscalars $(A_{1,2})$ are subsequently obtained through the 
following rotations~\footnote{We are implicitly assuming that all the parameters in the scalar potentials
are real so that the neutral scalars can be categorized into CP-odd and CP-even sectors respectively.}:
\begin{eqnarray}
\begin{pmatrix}  H_1^{\pm} \\ H_2^{\pm}\end{pmatrix} =
\begin{pmatrix}
 \cos\gamma_2 & -\sin\gamma_2 \\
 \sin\gamma_2 & \cos\gamma_2 \end{pmatrix} 
\begin{pmatrix} H_1^{\prime \pm} \\ H_2^{\prime \pm}\end{pmatrix} \label{e:chargedrot} \,,
	\quad
	\begin{pmatrix}  A_1 \\ A_2\end{pmatrix} =
	\begin{pmatrix}
            \cos\gamma_1 & -\sin\gamma_1 \\
		    \sin\gamma_1 & \cos\gamma_1 \end{pmatrix} 
	\begin{pmatrix}  A_1^{\prime} \\ A_2^{\prime}\end{pmatrix}  \label{e:pseudorot} \,.
\end{eqnarray}
The physical CP-even scalars, on the other hand, are directly obtained via a $3\times3$ rotation as\footnote{Note that, our convention of $\alpha_3$ differs from that in Ref.~\cite{Das:2019yad} by a minus sign.}
%\begin{subequations}
\begin{eqnarray}
	\begin{pmatrix}
		h \\ H_1 \\ H_2
	\end{pmatrix} = \begin{pmatrix}
	1 & 0 & 0 \\
	0 & \cos \alpha_3 &  -\sin \alpha_3  \\
	0 & \sin \alpha_3 & \cos \alpha_3 
	\end{pmatrix}
\begin{pmatrix}
	\cos \alpha_2 & 0 & \sin \alpha_2  \\
	0 & 1 & 0 \\
	-\sin \alpha_2 & 0 & \cos \alpha_2 
\end{pmatrix} \begin{pmatrix}
	\cos \alpha_1 & \sin \alpha_1 & 0 \\
	-\sin \alpha_1 & \cos \alpha_1 & 0 \\
	0 & 0 & 1 \end{pmatrix}
	\begin{pmatrix}
	h_\bl \\ h_\bd \\ h_\bu
\end{pmatrix} 	\label{e:cpevenrot}\,,
\end{eqnarray}
%\begin{eqnarray}
%	\begin{pmatrix}
%		h \\ H_1 \\ H_2
%	\end{pmatrix} = {\cal O}_\alpha
%	\begin{pmatrix}
%		h_\bl \\ h_\bd \\ h_\bu
%	\end{pmatrix} \,; \qquad {\cal O}_\alpha = {\cal R}_3 \cdot  {\cal R}_2\cdot {\cal R}_1 \,,	{\rm and,}	\label{e:cpevenrot}
%\end{eqnarray}
%%	\begin{eqnarray}\label{e:Oa}
%%		{\cal O}_\alpha &=& {\cal R}_3 \cdot  {\cal R}_2\cdot {\cal R}_1 \,,
%%	\end{eqnarray}
%%	%
%%	with
%%	%
%	\begin{equation}
%		\label{e:R}
%		%                                                                                                      
%	{\cal R}_1 = \begin{pmatrix}
%			\cos \alpha_1 & \sin \alpha_1 & 0 \\
%			-\sin \alpha_1 & \cos \alpha_1 & 0 \\
%			0 & 0 & 1 \end{pmatrix}
%		\,, {\cal R}_2 = \begin{pmatrix}
%			\cos \alpha_2 & 0 & \sin \alpha_2  \\
%			0 & 1 & 0 \\
%			-\sin \alpha_2 & 0 & \cos \alpha_2 
%		\end{pmatrix}\,,  
%		{\cal R}_3 = \begin{pmatrix}
%			1 & 0 & 0 \\
%			0 & \cos \alpha_3 &  -\sin \alpha_3  \\
%			0 & \sin \alpha_3 & \cos \alpha_3 
%		\end{pmatrix}\,,
%	\end{equation}
%\end{subequations}
where $h$ will be identified with the Higgs boson found at the LHC with mass $m_h = 125$ GeV. For later use,
it is useful to first define the CP-even states in the Higgs basis as 
\begin{eqnarray}
		\begin{pmatrix}
		H_{\rm SM} \\ H_1^\prime \\ H_2^\prime 
	\end{pmatrix} =
{\mathcal O}_\beta 
\begin{pmatrix}
	h_\bl \\ h_\bd \\ h_\bu 
\end{pmatrix}\,,
\end{eqnarray}
where $H_{\rm SM}$ is the state that possesses SM-like couplings at the tree-level. The 3HDM alignment limit is defined by~\cite{Das:2019yad}, 
\vspace{-0.3cm}
\begin{eqnarray}
	\alpha_1 = \beta_1 \quad {\rm and,}~ \alpha_2 = \beta_2\,.
	\label{e:alignment}
\end{eqnarray}
In the alignment limit, $H_{\rm SM}$ will coincide with the physical eigenstate $h$ and the remaining 
CP-even states $H_{1,2}$ will be related to $H_{1,2}^\prime$ as follows:
\begin{eqnarray}
		\begin{pmatrix}  H_1 \\ H_2\end{pmatrix} =
		\begin{pmatrix}
			\cos\alpha_3 & -\sin\alpha_3 \\
			\sin\alpha_3 & \cos\alpha_3 \end{pmatrix} 
		\begin{pmatrix} H_1^{\prime} \\ H_2^{\prime}\end{pmatrix} \label{e:h1h2rot} \,.
\end{eqnarray}

Now that we  have established our notation for the nonstandard scalars, we can now proceed
to write the Yukawa couplings for these scalars. It is convenient to write them first in terms
of $H_{1,2}^\prime$ and the $A_{1,2}^\prime$ in the neutral scalar sector as follows:
\begin{subequations}
\begin{eqnarray}
	- {\mathscr L}_Y^H &=& \dfrac{1}{v} \big[{\bar \bu} D_\bu \bu \left(H_{\rm SM} + \xi_{1\bu} H_1^\prime + \xi_{2\bu} H_2^\prime \right) + 
											{\bar \bd} D_\bd \bd \left(H_{\rm SM} + \xi_{1\bd} H_1^\prime + \xi_{2\bd} H_2^\prime \right)  \big. \nonumber \\
						&&					\big. +{\bar \bl} D_\bl \bl \left(H_{\rm SM} + \xi_{1\bl} H_1^\prime + \xi_{2\bl} H_2^\prime \right)\big]\,, \\
						\label{e:h1prime-yukawa} 
	- {\mathscr L}_Y^A &=& \dfrac{i}{v} \big[- {\bar \bu} D_\bu \gamma_5 \bu \left(\xi_{1\bu} A_1^\prime + \xi_{2\bu} A_2^\prime \right) + 
	{\bar \bd} D_\bd \gamma_5 \bd \left(\xi_{1\bd} A_1^\prime + \xi_{2\bd} A_2^\prime \right)  \big. \nonumber \\
	&&					\big. +{\bar \bl} D_\bl \gamma_5 \bl \left(\xi_{1\bl} A_1^\prime + \xi_{2\bl} A_2^\prime \right)\big]\,. 	\label{e:a1prime-yukawa}
\end{eqnarray}
\label{e:neutralprime-yukawa}
\end{subequations}
Here, $\bu, \bd$ and $\bl$ denote the physical up-type, down-type quarks and charged leptons respectively. 
The diagonal mass matrices $D_{\bu,\bd,\bl}$ are, therefore, given by,
\begin{eqnarray}
	D_\bu = {\rm diag} \{m_u, m_c, m_t\}\,, \quad D_\bd = {\rm diag} \{m_d, m_s, m_b\}\,, \quad D_\bl = {\rm diag} \{m_e, m_\mu, m_\tau\}\,.
	\label{e:diagmat}
\end{eqnarray}
The dimensionless coupling modifiers, denoted by $\xi$ in Eqs.(\ref{e:neutralprime-yukawa}) are given as:
\begin{subequations}
	\begin{eqnarray}
		\xi_{1\bu} = 0\,; && \quad \xi_{2\bu} = \dfrac{1}{\tan \beta_2} \,, \\
		\xi_{1\bd} = \dfrac{\sqrt{1 + \tan^2\beta_2}}{\tan \beta_1}\,; && \quad \xi_{2\bd} = -{\tan \beta_2} \,, \\
		\xi_{1\bl} = -\tan \beta_1\sqrt{1 + \tan^2\beta_2} \,; && \quad \xi_{2\bl} = -{\tan \beta_2} \,.
	\end{eqnarray}
	\label{e:xi-coup}
\end{subequations}
The Yukawa couplings of the neutral nonstandard physical scalars $H_{1,2}$ and $A_{1,2}$ can then be written as
\begin{subequations}
	\begin{eqnarray}
		- {\mathscr L}_Y^H &=& \dfrac{1}{v} \big[{\bar \bu} D_\bu \bu \left(h + \kappa_{\bu}^{H_1} H_1 + \kappa_{\bu}^{H_2} H_2 \right) + 
		{\bar \bd} D_\bd \bd \left(h + \kappa_{\bd}^{H_1} H_1 + \kappa_{\bd}^{H_2} H_2 \right) + \big. \nonumber \\
		&&					\big.{\bar \bl} D_\bl \bl \left(h + \kappa_{\bl}^{H_1} H_1 + \kappa_{\bl}^{H_2} H_2 \right)\big]\,, \\
		\label{e:cpeven-yukawa}
		- {\mathscr L}_Y^A &=& \dfrac{i}{v} \big[- {\bar \bu} D_\bu \gamma_5 \bu \left(\kappa_{\bu}^{A_1} A_1 + \kappa_{\bu}^{A_2} A_2 \right) + 
		{\bar \bd} D_\bd \gamma_5 \bd \left(\kappa_{\bd}^{A_1} A_1 + \kappa_{\bd}^{A_2} A_2 \right) + \big. \nonumber \\
		&&					\big.{\bar \bl} D_\bl \gamma_5 \bl \left(\kappa_{\bl}^{A_1} A_1 + \kappa_{\bl}^{A_2} A_2 \right)\big]\,. 	\label{e:cpodd-yukawa}
	\end{eqnarray}
	\label{e:neutral-yukawa}
\end{subequations}
The coupling modifiers for the physical nonstandard scalars can be obtained in a straightforward manner using the rotations 
in \Eqs{e:h1h2rot}{e:pseudorot} as follows:
\begin{subequations}
\begin{eqnarray}
	\kappa_f^{H_1} = \xi_{1f} \cos \alpha_3 - \xi_{2f} \sin\alpha_3 \,; && \quad \kappa_f^{H_2} = \xi_{1f} \sin \alpha_3 + \xi_{2f} \cos\alpha_3 \,, \label{e:kappah1h2}\\
	\kappa_f^{A_1} = \xi_{1f} \cos \gamma_1 - \xi_{2f} \sin\gamma_1 \,; && \quad \kappa_f^{A_2} = \xi_{1f} \sin \gamma_1 + \xi_{2f} \cos\gamma_1 \,, \label{e:kappaa1a2}
\end{eqnarray}\label{e:kappas}
\end{subequations}
where $f = \bu\,, \bd\,, \bl$. It should be noted that we have assumed the alignment limit to hold while writing \Eqn{e:kappah1h2}.
%%%%%%%%%%%%%%%%%%%%%%%%%%%%%%%%%%%%%%%%%%%%%%%%%%%%%

\subsection{The up-phobic limit}
\label{subsec:upphobic-limit}
\begin{figure}[t]
	\centering
	\includegraphics[width=80mm]{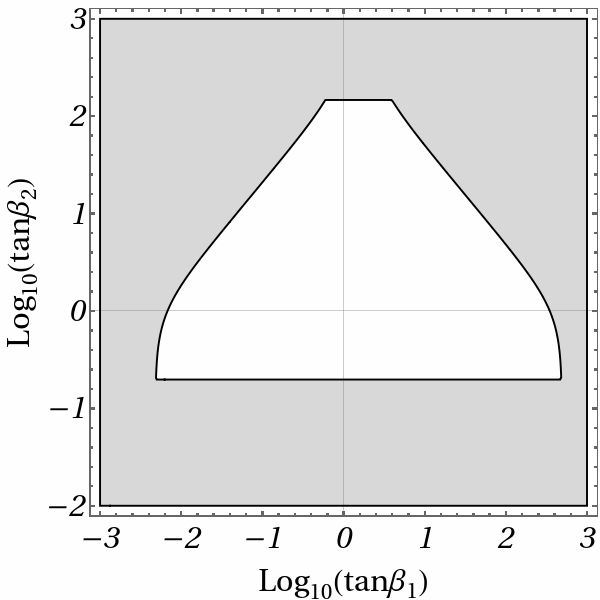}
	\caption{The constraints on the $\tan \beta_1$ vs $\tan\beta_2$ plane  from the requirement that all Yukawa couplings in the alignment and up-phobic limit are perturbative in nature. The gray shaded region is excluded.}
	\label{f:yukawa_perturb}
\end{figure}

Quite clearly, $H_1$ and $A_1$ will be up-phobic, {\it i.e.}, $H_1$ and $A_1$ would not couple to the up-type quarks at all in the limit,
\begin{subequations}
\begin{eqnarray}
	\kappa_\bu^{H_1(A_1)} = 0 \quad \Rightarrow \alpha_3 \,(\gamma_1) = 0\,.
	\label{e:upphobic-h1}
\end{eqnarray}
Similarly, the down-phobic and leptophobic limits of $H_1$ and $A_1$ should be obtained as
\begin{eqnarray}
 &&\kappa_\bd^{H_1\,(A_1)} = 0 \quad \Rightarrow \tan\alpha_3\,(\tan\gamma_1) = - \dfrac{1}{\tan \beta_1 \sin \beta_2}\,, \\
 && \kappa_\bl^{H_1\,(A_1)} = 0 \quad \Rightarrow \tan\alpha_3\,(\tan\gamma_1) = \dfrac{\tan\beta_1}{\sin \beta_2} \,. 
 \label{e:downphobic-leptophobic-h1}
\end{eqnarray} \label{e:limitsh1a1}
\end{subequations}
As complementary conditions, we can also obtain the limits in which $H_2$ and $A_2$ will be up-phobic, down-phobic and leptophobic
as follows:
\begin{subequations}
	\begin{eqnarray}
	 &&	\kappa_\bu^{H_2\,(A_2)} = 0 \quad \Rightarrow \alpha_3 \, (\gamma_1) = \dfrac{\pi}{2}\,, \\
		&&\kappa_\bd^{H_2\,(A_2)} = 0 \quad \Rightarrow \tan\alpha_3 \, (\tan\gamma_1) = {\tan \beta_1 \sin \beta_2}\,, \\
		&& \kappa_\bl^{H_2\,(A_2)} = 0 \quad \Rightarrow \tan\alpha_3 \,( \tan\gamma_1 )= - \dfrac{\sin \beta_2}{\tan\beta_1} \,. 
	\end{eqnarray} \label{e:limitsh2a2}
\end{subequations}
%
%%%%%%%%%%%%%%%%Figures%%%%%%%%%%%%%%%%%%%%%%%%%%%%%%%%%%%%%%%%%
\begin{figure}[ht]
	\centering
	\includegraphics[width=80mm]{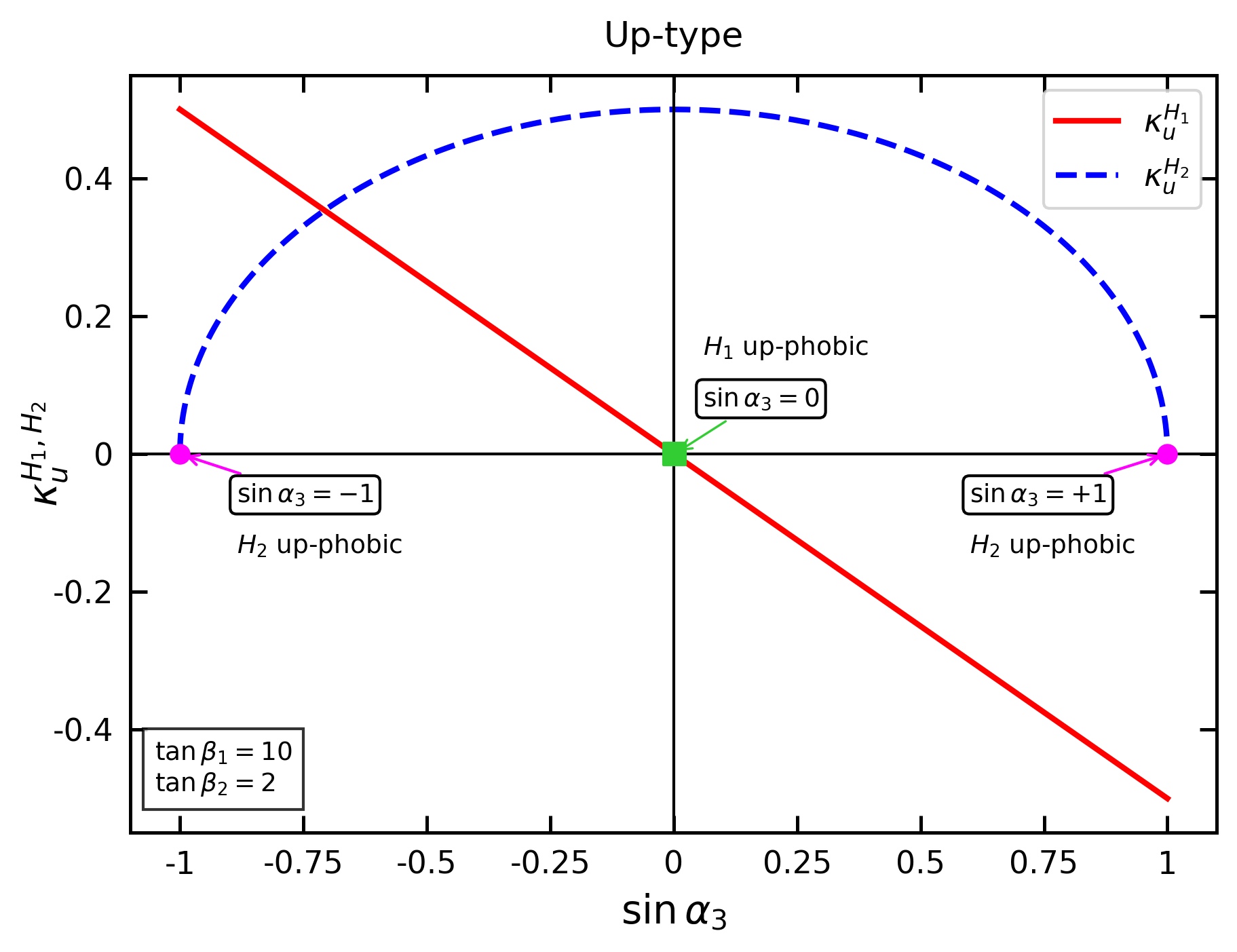} 
	\includegraphics[width=80mm]{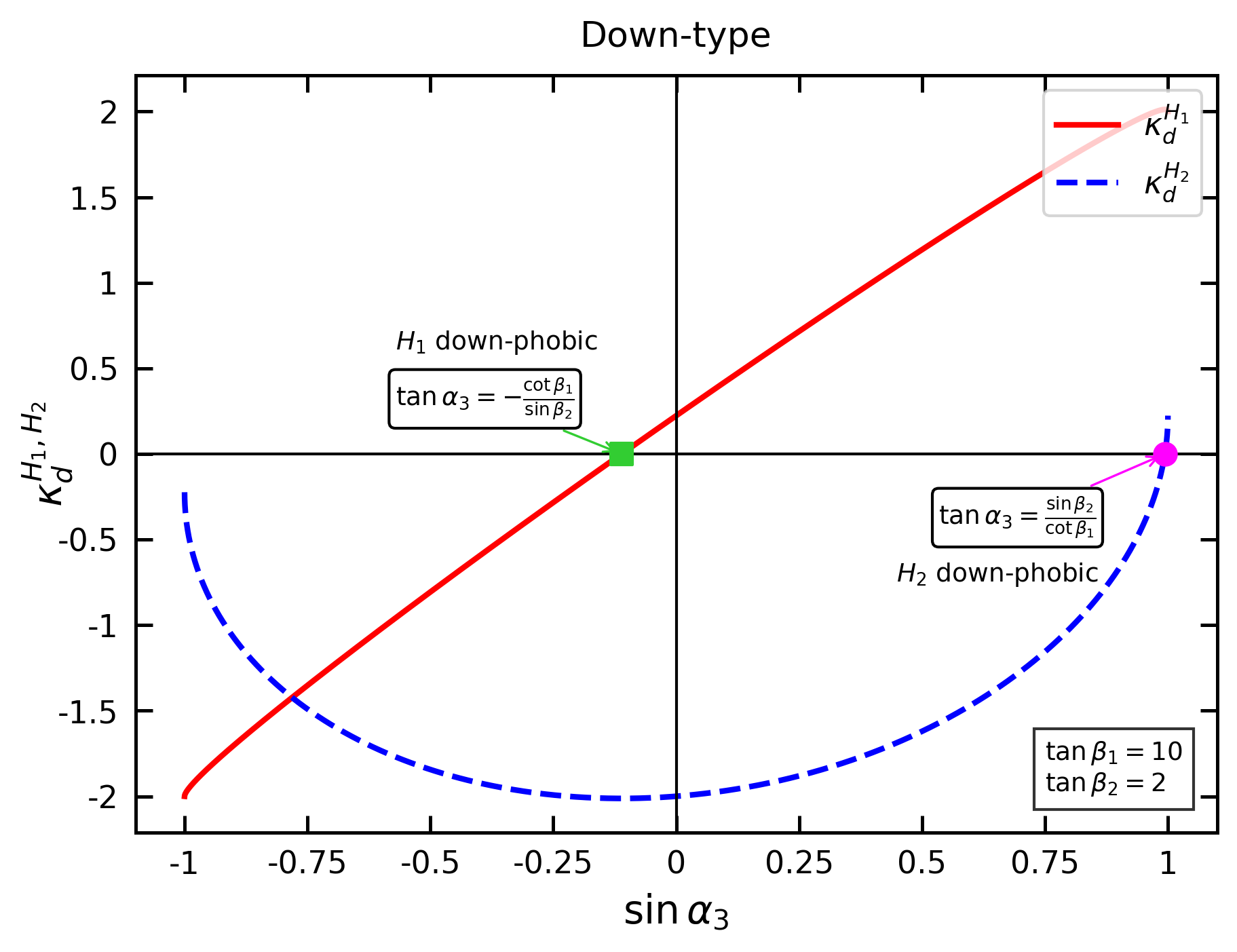}
	\includegraphics[width=80mm]{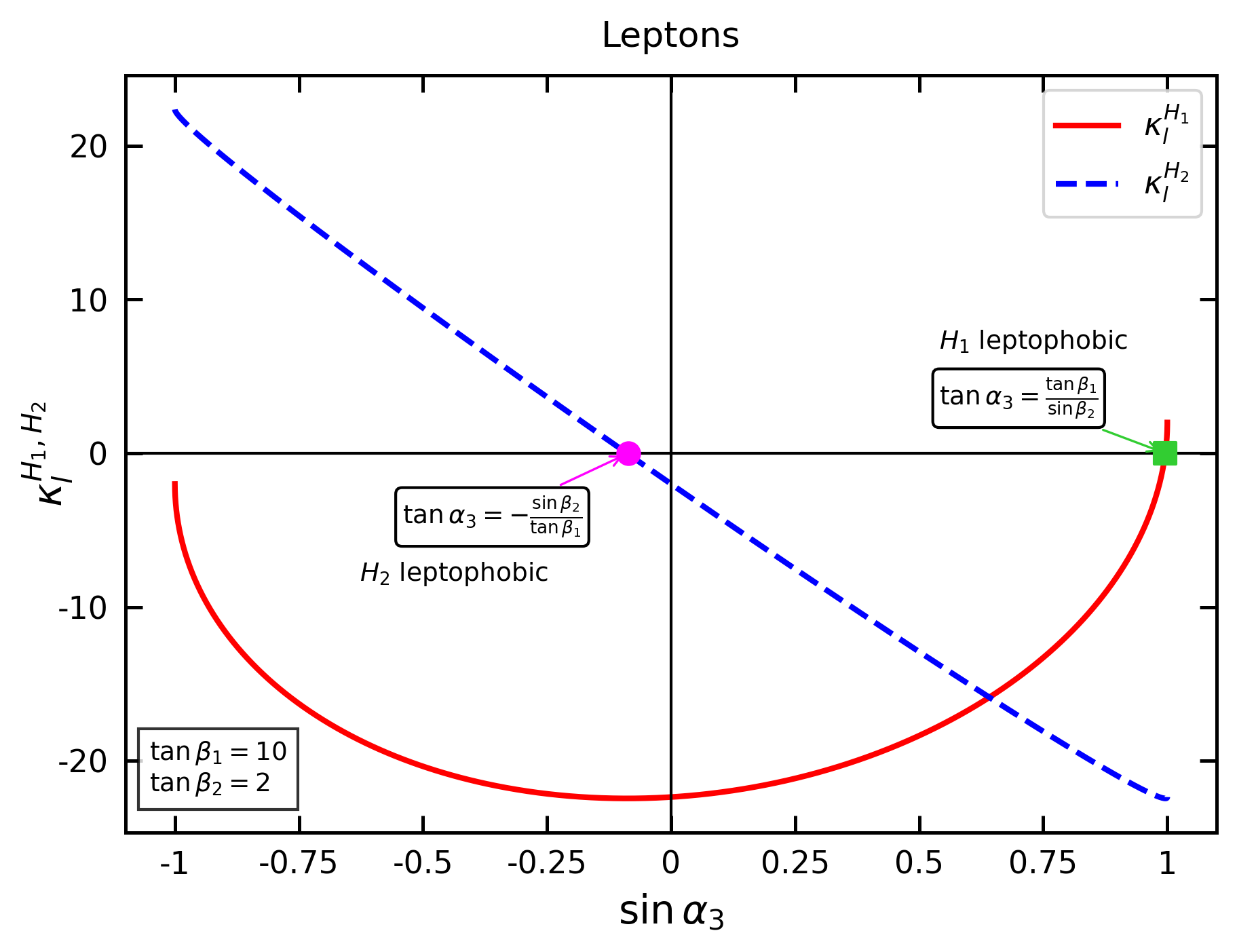}
	\caption{The up-phobic, down-phobic and leptophobic limits of $H_{1,2}$ for a benchmark point of $\tan\beta_1 = 10$
		and $\tan\beta_2 = 2$. The corresponding limits for $A_{1,2}$ will exhibit the same pattern with $\alpha_3$ replaced by 
		$\gamma_1$.}
	\label{f:limits}
\end{figure}
These limits are summarized in tabular form in Table~\ref{tab:phobic_limits}.
In Fig.~\ref{f:limits}, we have displayed these limits of \Eqs{e:limitsh1a1}{e:limitsh2a2} explicitly with a benchmark choice of
$\tan\beta_1 = 10$ and $\tan\beta_2= 2$.  
In Fig.~\ref{f:yukawa_perturb}, we show the allowed values of $\tan\beta_{1,2}$ from the requirement that the Yukawa couplings are perturbative in nature which essentially restricts the $\xi_f^{1,2}$s, as can be understood from \Eqs{e:neutral-yukawa}{e:kappas}. Consequently, the gray shaded region is excluded from the requirements~\cite{Boto:2024jgj}
	\begin{eqnarray}
		\left|\xi_f^{1,2} \dfrac{m_f}{v}\right| < \sqrt{4\pi}\,, \quad (f = \bu\,,\bd,\bl)\,. \label{e:yukawa-perturbativity}
	\end{eqnarray} 
	Accordingly, we choose our benchmarks of $\tan\beta_{1,2}$ throughout this paper within this limit and mark the disallowed region in other cases. It is important to highlight that, in the NFC versions of the 2HDM, the coupling modifiers defined in the same way as in 
\Eqs{e:limitsh1a1}{e:limitsh2a2}, can at most be suppressed. In contrast, in a Type-$Z$ or democratic 3HDM,
such coupling modifiers can vanish identically, rendering the corresponding neutral scalars to be truly
up-phobic, down-phobic or leptophobic as the case may be.

For completeness, we also display the charged Higgs Yukawa Lagrangian as follows:
\begin{eqnarray}
	- {\mathscr L}_Y^C &=& \dfrac{\sqrt{2}}{v} H_1^{\prime + } \big[{\bar \bu} \left(V D_\bd \xi_{1\bd} P_R - D_\bu V \xi_{1\bu} P_L\right)  \bd   + 
	\bm{\bar \nu} \left(D_\bl\xi_{1\bl} P_R\right)  \bl \big] + \nonumber \\
	&&	\dfrac{\sqrt{2}}{v} H_2^{\prime + } \big[{\bar \bu} \left(V D_\bd \xi_{2\bd} P_R - D_\bu V \xi_{2\bu} P_L\right)  \bd   + 
	\bm{\bar \nu} \left(D_\bl\xi_{2\bl} P_R\right)  \bl \big] + {\rm h.c.} \,,
	\label{e:charged-yukawa}
\end{eqnarray}
where $V$ is the CKM matrix. The physical charged Higgs coupling can be obtained from this by applying the rotation in \Eqn{e:chargedrot}.
%%%%%%%%%%%%%%%%%%%%%%%%%%%%%%%%%%%%%%%%%%%%%%%%%%%%%%%%%%%%%%%%%%%%%%%%%%%%%%%%%%%%%%%%%%%

%%%%%%%%%%%%%%%%%%%%%%%%% Table %%%%%%%%%%%%%%%%%%%%%%%%%%%%%%%%%%%%%%%%%%%%%%%%%

%---------------------------------------------------------
% Table
%---------------------------------------------------------
\begin{table}[t]
	\centering
	\renewcommand{\arraystretch}{2}
	
	\begin{tabular}{l|c|c}
		\toprule
		\rowcolor{gray!15}
		\textbf{Condition} &
		\textbf{$S=H_1(A_1)$} &
		\textbf{$S=H_2(A_2)$} \\
		\midrule
		
		$\mathbf{\kappa}_\bu^S=0$ (Up-phobic)
		&
		$\displaystyle \alpha_3\,(\gamma_1)=0$
		&
		$\displaystyle \alpha_3\,(\gamma_1)=\frac{\pi}{2}$
	    \\[2mm] \hline
		
		$\mathbf{\kappa}_\bd^S=0$ (Down-phobic)
		&
		$\displaystyle
		\tan\alpha_3\,(\tan\gamma_1)=
		-\frac{1}{\tan\beta_1\,\sin\beta_2}$
		&
		$\displaystyle
		\tan\alpha_3\,(\tan\gamma_1)=
		\tan\beta_1\,\sin\beta_2$
		\\[3mm] \hline
		
		$\mathbf{\kappa}_\bl^S=0$ (Leptophobic)
		&
		$\displaystyle
		\tan\alpha_3\,(\tan\gamma_1)=
		\frac{\tan\beta_1}{\sin\beta_2}$
		&
		$\displaystyle
		\tan\alpha_3\,(\tan\gamma_1)=
		-\frac{\sin\beta_2}{\tan\beta_1}$
		\\
		
		\bottomrule
	\end{tabular}	
	\caption{Conditions corresponding to the up-phobic, down-phobic and leptophobic limits of the neutral nonstandard Higgs bosons. 
	Definitions of $\kappa$s are given in \Eqn{e:kappas}.}
	\label{tab:phobic_limits}	
\end{table}
%%%%%%%%%%%%%%%%%%%%%%%%%%%%%%%%%%%%%%%%%%%%%%%%%%%%%%%%%%%%%%%%%%%%%%%%%%%%%%%%%%%5
%%%%%%%%%%%%%%%%%%%%%%%%%%%%%%%%%%%%%%%%%%%%%%%%%%%%%

\section{Constraints and phenomenological implications}
\label{sec:pheno}
For the phenomenological analysis, we will work in the `theoretically safe' limit~\cite{Darvishi:2021txa,Chakraborti:2021bpy},
\begin{eqnarray}
	m_{H_1} = m_{A_1} = m_{H_1^\pm} = M_1 \,; \quad m_{H_2} = m_{A_2} = m_{H_2^\pm} = M_2\,; \quad \gamma_1 = \gamma_2 = \alpha_3 = \alpha\,,
	\label{e:theolimit}
\end{eqnarray}
where, $M_1$, $M_2$ and $\alpha$ have been defined in \Eqn{e:soft terms scale} in the appendix. We will also assume the alignment limit defined by \Eqn{e:alignment} to hold so that the constraints from Higgs signal
	strength measurements are also fulfilled.  As we argue after \Eqn{e:potential_sp6} in the appendix, the constraints arising from unitarity and BFB~\cite{Bento:2017eti,Boto:2021qgu,Bento:2022vsb,Boto:2022uwv,Jurciukonis:2026dmc}
	are automatically satisfied in this limit. From the results of Ref.~\cite{Das:2022gbm}, we can conclude that the constraints from the 
	electroweak $T$-parameter will also be obeyed in this limit.
We will focus particularly in the up-phobic limit of $H_1$ and $A_1$, i.e., we set $\alpha = 0$.

For the numerical analysis, we implemented the model in FeynRules-v2.3~\cite{Alloul:2013bka}, and the corresponding UFO model files~\cite{Degrande:2011ua} have been generated for use with MadGraph5-v3.6.3~\cite{Alwall:2014hca}. Using this implementation, the production cross sections and branching ratios of the scalar particles have been computed and subsequently employed for the evaluation of the current direct search limit. .

\begin{figure}[ht]
	\centering
	\includegraphics[height=70mm,width=70mm]{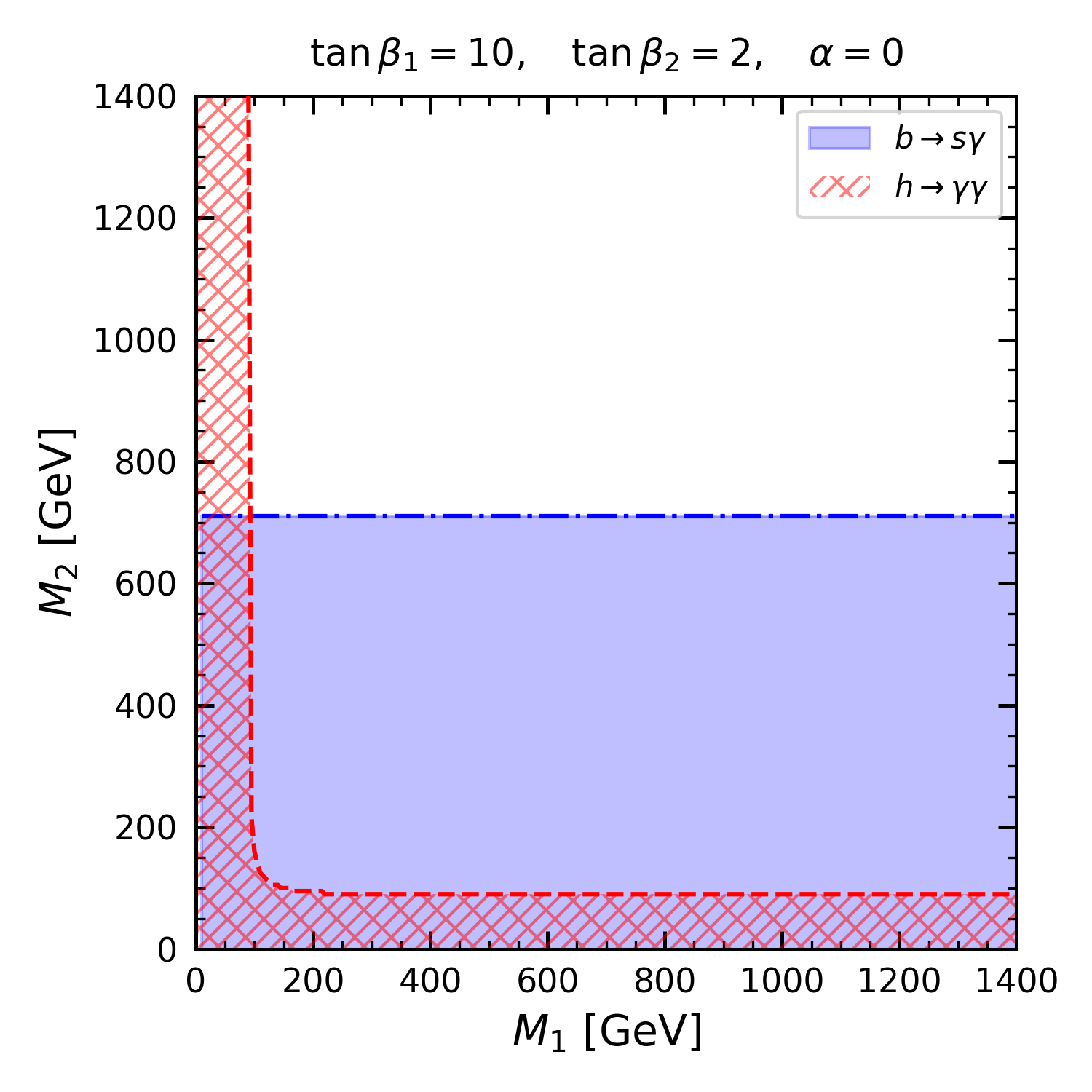} 
	\includegraphics[height=70mm,width=70mm]{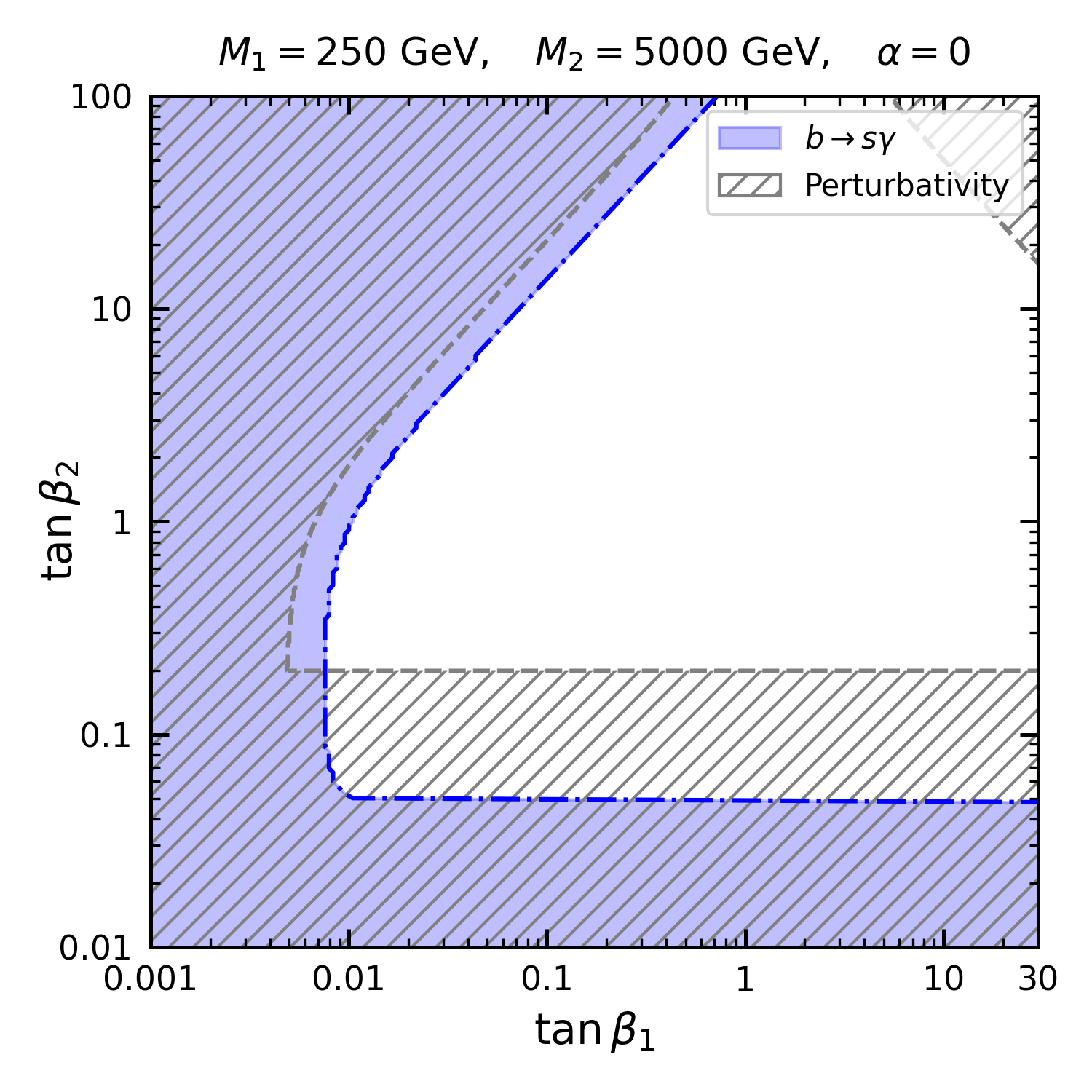} \\
	\caption{Current constraints in the $M_1$-$M_2$ plane (left) and $\tan\beta_1$-$\tan\beta_2$ plane (right) from the $b\to s\gamma$ observable~\cite{ParticleDataGroup:2026aaa}. The blue shaded region is excluded at 95\% C.L. The red hatched region in the left panel denotes the excluded parameter space from the current Higgs to diphoton signal strength, $\mu_{\gamma\gamma}$ at 2$\sigma$~\cite{ATLAS:2026pdi}. The gray hatched region in the right panel is excluded from the requirement of perturbativity of the Yukawa couplings.
	}
	\label{f:flavor}
\end{figure}

Before turning to the LHC phenomenology, we first discuss the constraints from flavor observables and the Higgs diphoton signal strength, summarized in Fig.~\ref{f:flavor}. Following the procedure of Ref.~\cite{Chakraborti:2021bpy}, we derive the flavor constraints and, in particular, display the current 95\% C.L. bound on $M_2$ arising from $b \to s \gamma$~\cite{ParticleDataGroup:2026aaa}. A notable feature emerges in the up-phobic limit, $\alpha = 0$, where the flavor constraints on $M_1$ are substantially weakened. In particular, for our chosen benchmark of $\tan\beta_{1,2}$ the $b \to s\gamma$ bound on $M_2$ in the up-phobic limit becomes essentially independent of $M_1$, as shown in the left panel of Fig.~\ref{f:flavor}, where the shaded blue region indicates the excluded parameter space. Furthermore, for $M_2$ in the TeV range, no constraint applies to the region with $\tan\beta_{1,2}>1$ even for a light $M_1$ at the electroweak scale, as shown in the right panel of Fig.~\ref{f:flavor}, where the blue shaded region again marks the disallowed parameter space. In the right panel, the parameter region excluded by the requirement of perturbativity of the Yukawa couplings is indicated by the gray hatched area.

Moreover, the analysis of Ref.~\cite{Chakraborti:2021bpy} shows that constraints from neutral-meson mass differences, $\Delta M$, do not significantly restrict $M_1$ in the up-phobic regime, particularly for $\tan\beta_2\gtrsim 1$. The flavor observables therefore comfortably accommodate a first tier of non-standard scalars at the electroweak scale, while the second tier remains at $\mathcal{O}(\text{TeV})$ masses.

We next turn to constraints from the loop-induced decay $h \to \gamma\gamma$, which receives additional contributions from the charged Higgs bosons in the model. In the alignment limit together with the limit of \Eqn{e:theolimit}, the relevant $hH_{1,2}^{+}H_{1,2}^-$ couplings take a particularly simple form, independent of $\tan\beta_{1,2}$, as 
	can be seen from \Eqn{e:ghch-theolimit} in the appendix.
%\sout{; explicit expressions are given in Ref.~\mbox{\cite{Chakraborti:2021bpy}}}. 
The diphoton signal strength thus provides a complementary constraint on the charged-scalar masses. In the left panel of Fig.~\ref{f:flavor}, the red hatched region is excluded at $2\sigma$ by the latest measurement, $\mu_{\gamma\gamma} = 1.03^{+0.09}_{-0.08}$~\cite{ATLAS:2026pdi}. This constraint is also sufficiently mild that an electroweak-scale $M_1$ remains viable.

Taken together, the flavor and Higgs signal-strength data therefore permit a spectrum in which $M_1$ lies at the electroweak scale while $M_2$ is of ${\cal O}(\rm TeV)$, particularly in the up-phobic limit of $H_1,A_1$. This provides a consistent phenomenological basis for the benchmark scenarios considered in the remainder of this work. We consequently focus on the LHC phenomenology of the Type-$Z$ (democratic) 3HDM in this limit, where the unconventional collider phenomenologies associated with the electroweak-scale first tier of non-standard scalars can be investigated through direct searches at the LHC. 

\begin{figure}[t]
	\centering
	\includegraphics[height=50mm,width=55mm]{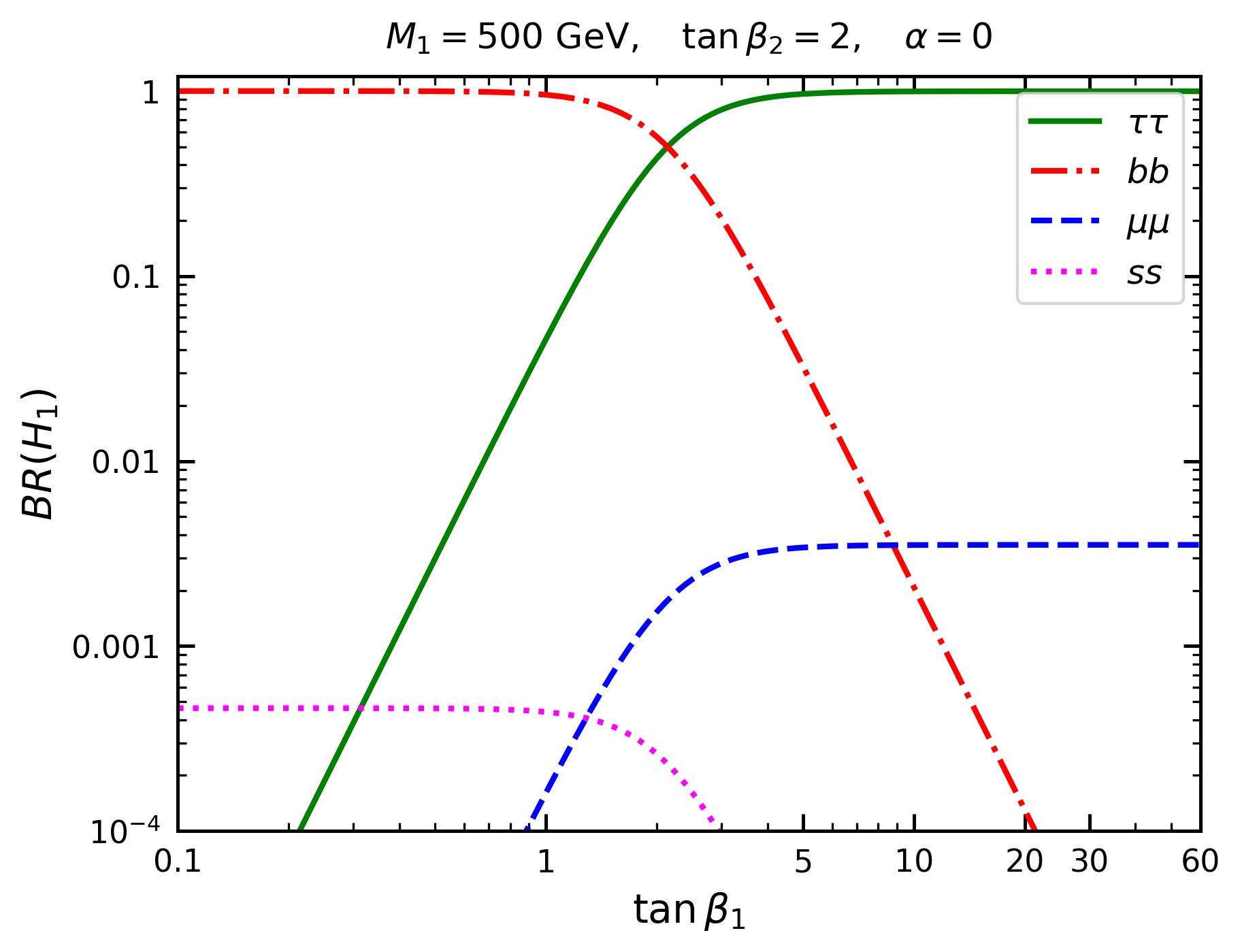} 
	\includegraphics[height=50mm,width=55mm]{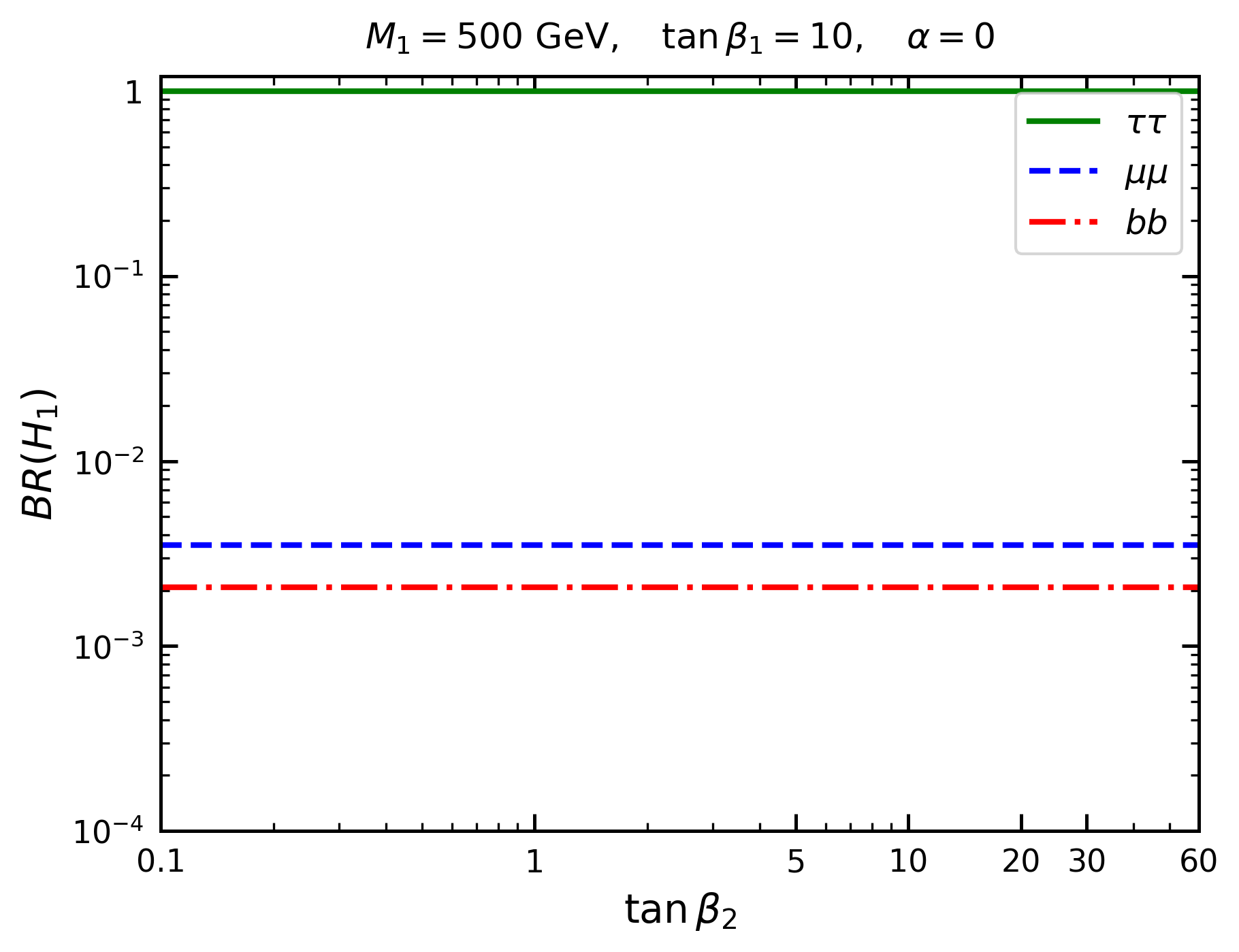}
	\includegraphics[height=50mm,width=55mm]{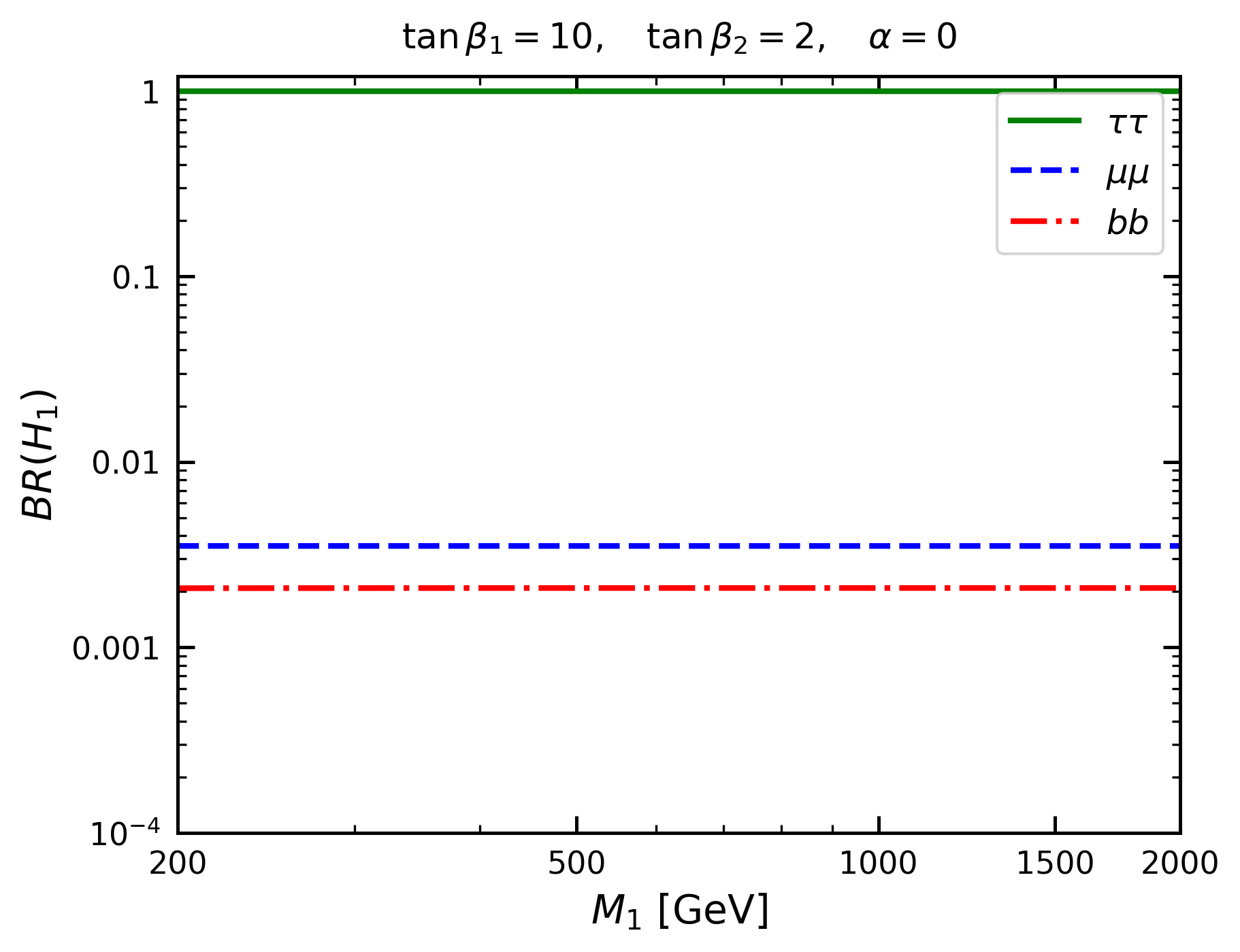} \\
	\includegraphics[height=50mm,width=55mm]{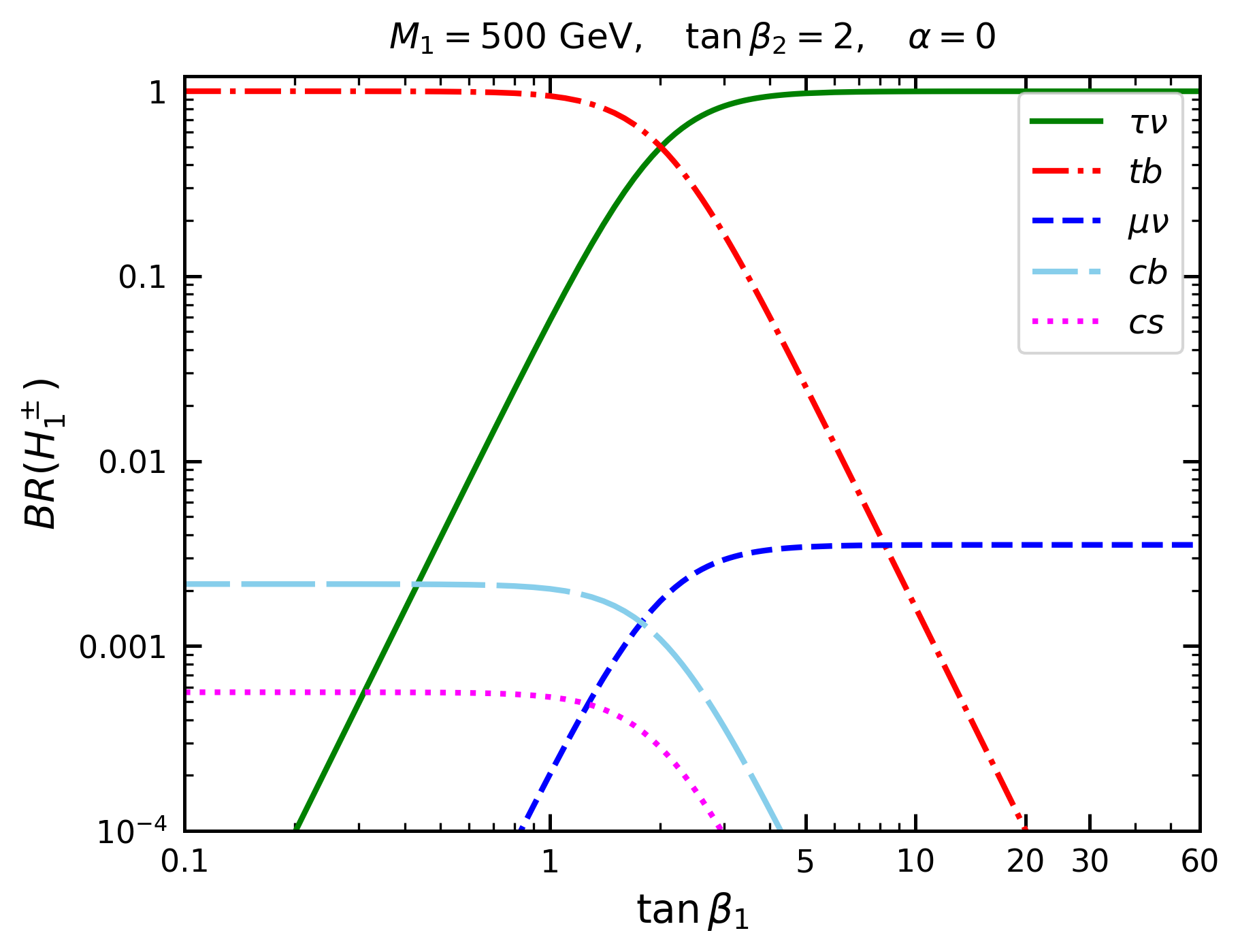} 
	\includegraphics[height=50mm,width=55mm]{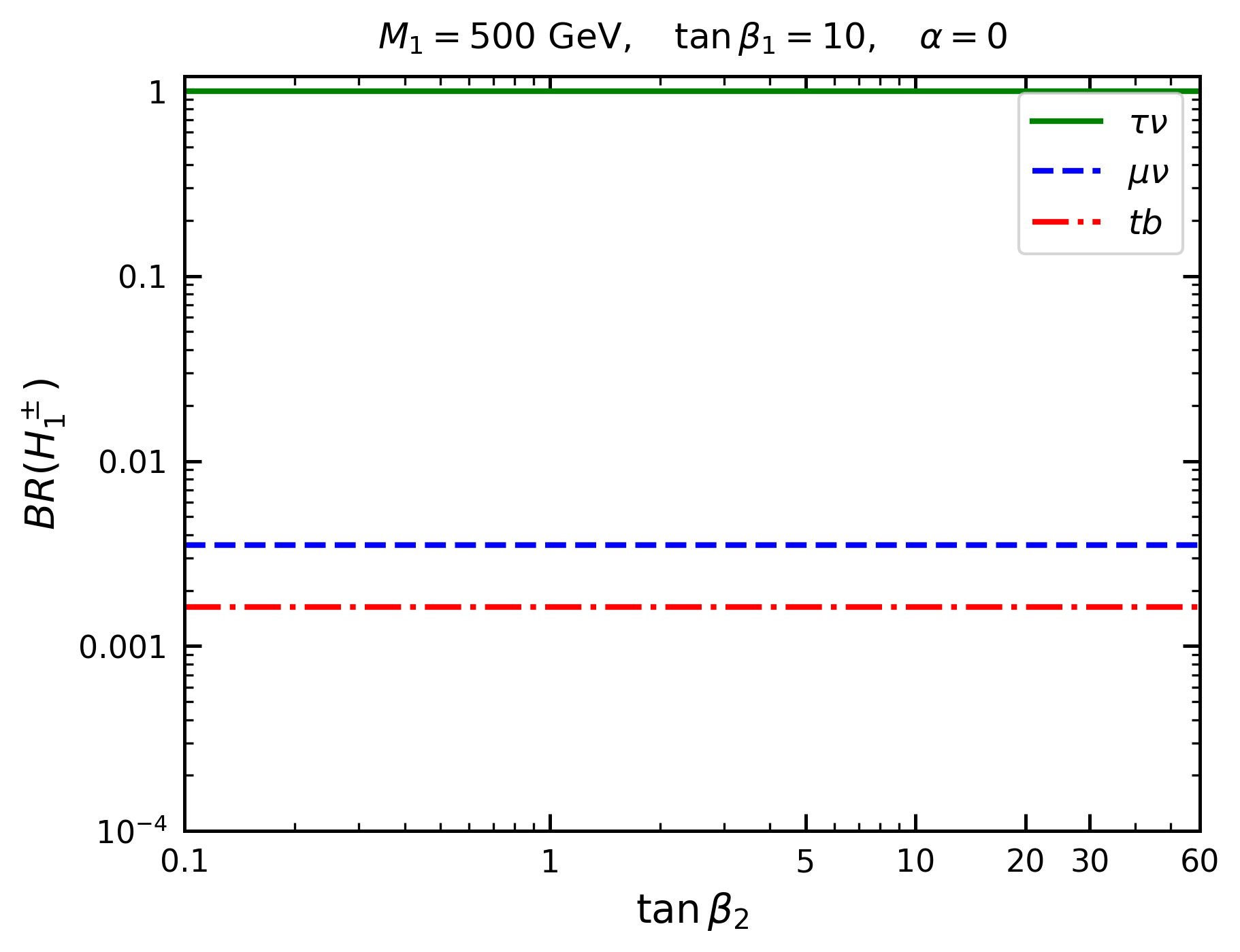}
	\includegraphics[height=50mm,width=55mm]{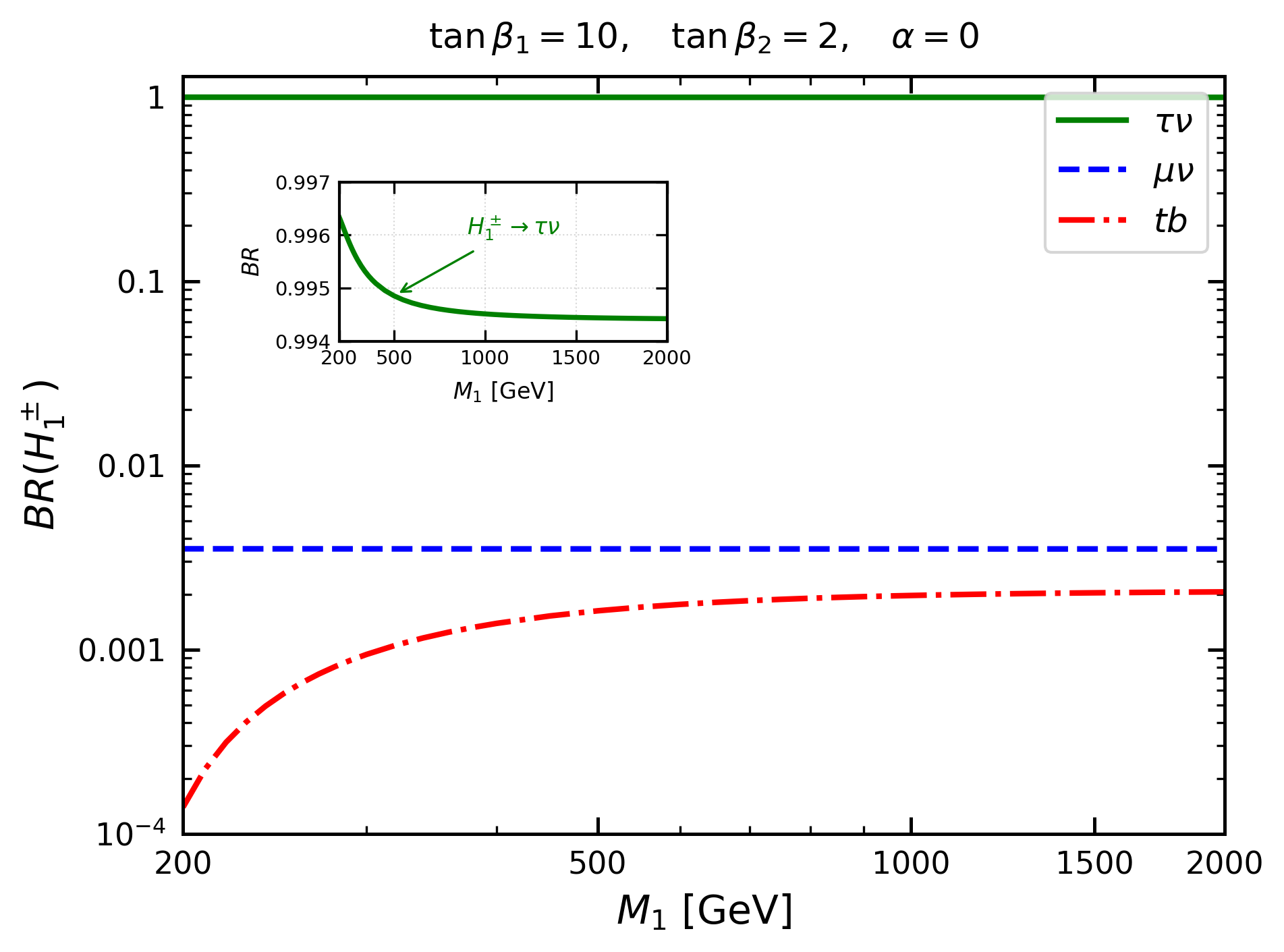} \\
	\caption{Branching ratios of $H_1$(BRs of $A_1$ are similar in nature) and $H_1^\pm$ for varying 
		$\tan\beta_1\,, \tan\beta_2\,,$ and the common mass $M_1$ in the up-phobic limit. The inset in the lower right
		panel shows a zoomed in view of the $H_1^\pm \to \tau \nu$ branching ratio. The corresponding feature is not visible in the main plot due to its small variation. A similar feature is also observed in the $H_1^\pm \to \mu \nu$ mode.  
	}
	\label{f:BR}
\end{figure}

To assess the relevance of $b$-associated production of the neutral scalars in the up-phobic limit, we first evaluate the branching ratios (BRs) of the neutral scalar $H_1$ (and $A_1$) and the charged scalar $H_1^\pm$, as shown in Fig.~\ref{f:BR}. In the three panels, we study the dependence of the branching ratios on $\tan\beta_1$, $\tan\beta_2$, and the common mass $M_1$, respectively, while keeping the remaining parameters fixed. The values of the other parameters which are kept fixed at some benchmark values are indicated in the corresponding figure legends. We further assume $M_1\ll M_2$ and fix $M_2\simeq 5$ TeV, such that the second tier of heavy nonstandard Higgses does not affect the phenomenology at the $\mathcal{O}(\rm 1~TeV)$ scale.

For fixed $\tan\beta_1$, the relative BRs of $H_1$ ($A_1$) exhibit no dependence on $\tan\beta_2$ owing to the structure of the fermionic couplings in the up-phobic limit, given in \Eqn{e:xi-coup}. Note that, both the down-type quark and charged-lepton Yukawa couplings of $H_1$ ($A_1$) has same $\tan\beta_2$ dependence which makes the relative BRs effectively independent of $\tan\beta_2$. Moreover, for the neutral scalars, the absence of the top-quark decay mode in the up-phobic limit renders their BRs largely insensitive to $M_1$ as well. It is worth noting that, in the alignment limit, decay modes such as $A_1\to Zh$, $H_1^\pm\to W^\pm h$ and $H_1 \to h h$ are absent.

Quite remarkably, in a very wide region of the 3HDM parameter space, we obtain the decay hierarchy,
\begin{eqnarray}
{\rm BR}(H_1, A_1 \to \tau^+ \tau^-) > {\rm BR}(H_1\,, A_1 \to b \bar b) \gg {\rm BR}(H_1\,, A_1 \to t \bar t). \label{e:H1-decay-hierarchy}
\end{eqnarray}
We wish to elaborate further on this point. Consider first a neutral Higgs boson with flavor-universal Yukawa couplings (modulo the fermion mass) as in the SM,
\begin{eqnarray}
	{\cal L}^h_{\rm Yuk} = - \dfrac{m_f}{v} \bar f f h,
	\qquad (f = {\bf u, d, \bl}).
	\label{e:higgs-yukawa}
\end{eqnarray}
Since $m_t\gg m_b>m_\tau$, one naturally expects
\begin{eqnarray}
	{\rm BR}(h\to t\bar t)\gg {\rm BR}(h\to b\bar b) >
	{\rm BR}(h\to\tau^+\tau^-)
	\label{e:h-decay-hierarchy}
\end{eqnarray}
when all three channels are kinematically open. A significant departure from this `mass-driven' relative hierarchy therefore points to a Higgs boson whose couplings can distinguish among the different fermion sectors.

Such a distinction is limited in a NFC based 2HDM variant. For example, consider the Type-II 2HDM where the up-type quarks couple to one Higgs-doublet (usually labeled as $\phi_2$), while down-type quarks and charged leptons both couple to the other (usually denoted by $\phi_1$). Notably, the down-type quarks and charged leptons are still treated on the same footing. Consequently,
\begin{eqnarray}
	{\rm BR}(S^0\to b\bar b) > {\rm BR}(S^0\to\tau^+\tau^-)
	\label{e:H-decay-hierarchy}
\end{eqnarray}
is still maintained and thus, the hierarchy in Eq.~(\ref{e:h-decay-hierarchy}) cannot be completely reversed~\cite{Kanemura:2014dea,Aiko:2020ksl}.  This follows from the fact that two scalar doublets can distinguish at most two of the three massive fermion sectors.
In contrast, the democratic 3HDM assigns separate scalar doublets to up-type quarks, down-type quarks, and charged leptons. Accordingly, in a sizable region of parameter space in Fig.~\ref{f:BR}, we find the hierarchy as in \Eqn{e:H1-decay-hierarchy}
representing a complete reversal of Eq.~(\ref{e:h-decay-hierarchy}). Observation of such a hierarchy would therefore provide a characteristic signature distinguishing the democratic 3HDM from its 2HDM counterparts.

The generality of this hierarchy within the democratic 3HDM framework can be understood from the corresponding Yukawa couplings. 
	Note that, for Eq.~(\ref{e:H1-decay-hierarchy}) to hold one would only require the $H_1 \tau^+ \tau^-$ Yukawa coupling
	to overshoot the $H_1 b \bar b$ Yukawa coupling which in turn needs to be much larger than the $H_1 t \bar t$ Yukawa coupling. This, in
	essence, would require (from \Eqn{e:neutral-yukawa})
\begin{eqnarray}
	\kappa_\bu^{H_1}(\alpha_3,\beta_1,\beta_2)m_t
	\ll
	\kappa_\bd^{H_1}(\alpha_3,\beta_1,\beta_2)m_b <
	\kappa_\bl^{H_1}(\alpha_3,\beta_1,\beta_2)m_\tau,
	\label{e:H1-kappa}
\end{eqnarray}
so that the hierarchy is controlled only by $\alpha_3$, $\tan\beta_1$, and $\tan\beta_2$. Once these are specified, the additional assumptions, such as \Eqn{e:theolimit}, do not affect the relative hierarchy of the $H_1$ Yukawa couplings or the hierarchies among its fermionic branching fractions. Since $\alpha_3$ is fixed by the up-phobic limit, the hierarchy is ultimately controlled by $\tan\beta_1$ and $\tan\beta_2$.

As evident from Fig~\ref{f:BR}, for $\tan\beta_1>2$, the neutral scalars predominantly decay into $\tau^+\tau^-$, whereas for $\tan\beta_1\lesssim2$, the $b\bar b$ mode becomes dominant, making $b$-associated production followed by $\tau^+\tau^-$ decay the most promising search channel.
In the lower panel of Fig.~\ref{f:BR}, we show the BRs of the charged scalar $H_1^\pm$. 
As we can see, for fixed $\tan\beta_2 = 2$, the dominant decay mode of $H_1^+$ is $\tau^+\nu$ for large values of $\tan\beta_1 \gg 1$ , while 
the $t \bar b$ mode dominates at small $\tan\beta_1 \lesssim 2$. Therefore, we also considered the current constraints from charged-scalar searches in the $t\bar b$ production channel with subsequent decays $H_1^+\to\tau^+\nu$ and $H_1^+\to t\bar b$ for completeness. We discuss our findings in the following paragraphs.

\begin{figure}[ht]
	\centering
	\includegraphics[width=80mm]{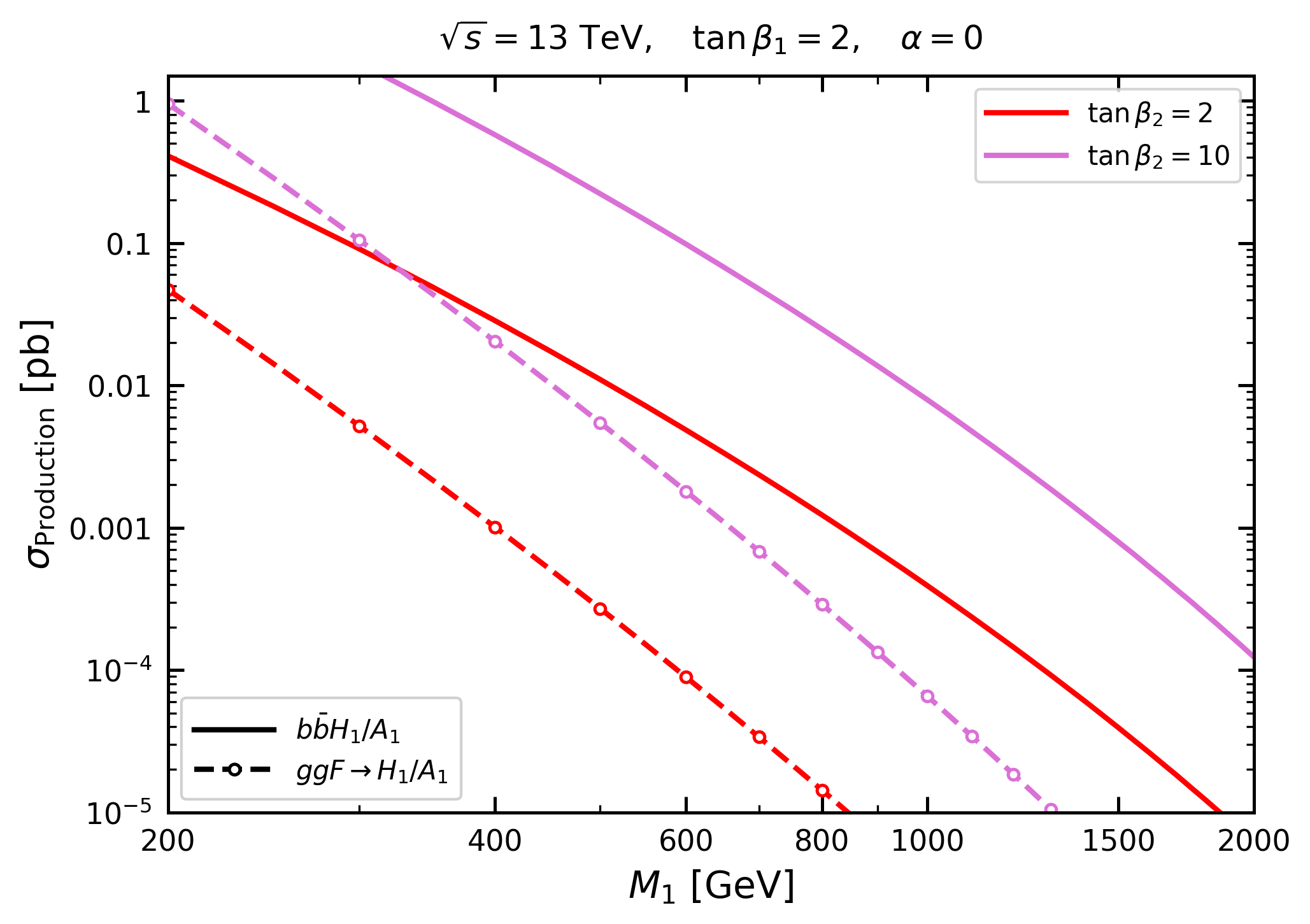} 
	\includegraphics[width=80mm]{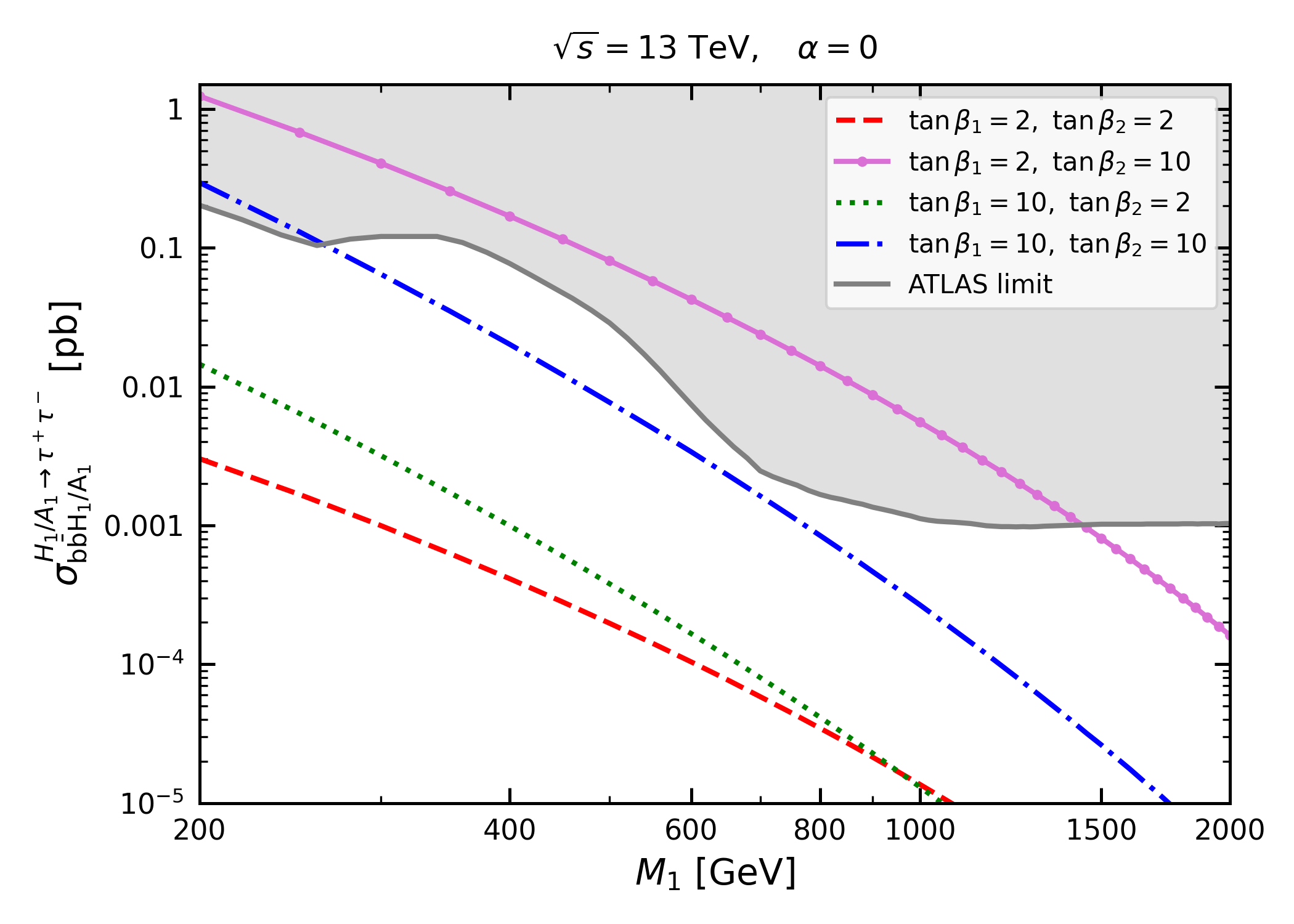} 
	\caption{(Left) The total production cross-section for heavy Higgs bosons in the gluon-gluon fusion (ggF) channel $(pp \to H_1/A_1, H_1/A_1 \to \tau^+ \tau^-)$ (dashed), together with $b$-associated production channel $(pp \to b \bar{b} H_1/A_1, H_1/A_1 \to \tau^+ \tau^-)$ (solid), for two benchmark values of $\tan \beta_2$ with $\tan\beta_{1}=2$, in the up-phobic limit. (Right) The effective cross section for $(pp \to b\bar b H_1/A_1, H_1/A_1 \to \tau^+ \tau^-)$ is shown for several benchmark values of $\tan\beta_1$ and $\tan\beta_2$ in the up-phobic limit. The gray shaded region indicates the current ATLAS limit on the effective cross-section at $13$ TeV and $139$ $fb^{-1}$ integrated luminosity~\cite{ATLAS:2020zms}.}
	\label{f:xsec}
\end{figure}

	 %In Fig.~\ref{f:xsec}, we display the cross-section for the $b$-associated production of the neutral scalars, $pp\to b\bar b H_1/A_1$  at the LHC with a centre-of-mass energy of $13~\mathrm{TeV}$ in the up-phobic limit
	 %for two benchmark points of $\tan\beta_2$ keeping $\tan\beta_1 = 2$ fixed.  
	 To directly compare the gluon-fusion (ggF) and $b$-associated production modes for the neutral scalars, we present in the left panel of Fig.~\ref{f:xsec} their total production cross sections at $\sqrt{s}=13~\mathrm{TeV}$ in the up-phobic limit for two benchmark choices of $\tan\beta_{1,2}$. The solid and dashed curves correspond to $b$-associated production and ggF, respectively. As expected, ggF is suppressed in this limit, with its cross section being approximately an order of magnitude smaller than that of the $b$-associated channel for the considered values of $\tan\beta_1$ and $\tan\beta_2$. Consequently, we focus on the $b$ associated production mode in the following analysis and do not present results for ggF channel.
	 %
	 %In the left panel, we present the total production cross section for different combinations of $\tan\beta_1$ and $\tan\beta_2$. 
	 In the right panel, we show the current ATLAS limits~\cite{ATLAS:2020zms} on the effective $b$-associated cross section where the neutral scalar decay to $\tau^+\tau^-$ mode $(pp \to b \bar b H_1/A_1, H_1/A_1 \to \tau^+ \tau^-)$ at $\sqrt{s}=13~\mathrm{TeV}$ with an integrated luminosity of $139~\mathrm{fb}^{-1}$. The gray shaded region is excluded from the current limit. %for the process
	 The different curves represent the corresponding model predictions for different sets of $\tan\beta_1$ and $\tan\beta_2$ values.

%	 In the right panel of Fig.~\ref{f:xsec_ggf}, we show the effective cross section for the process $pp\to H_1/A_1\to\tau^+\tau^-$, considering the ggF production mode only, for the same benchmark points. For comparison, we also display the current ATLAS constraint on the corresponding effective cross section at $\sqrt{s}=13~\mathrm{TeV}$ with an integrated luminosity of $139~\mathrm{fb}^{-1}$, with the gray shaded region indicating the excluded parameter space. Most of the benchmark points remain within the allowed region. 

%%%%%%%%%%%%%%%%%%%% Xsec &  BR figure%%%%%%%%%%%%%%%%%%%%%%%%%%%%%%%%%%%
%

%%%%%%%%%%%%%%%%%%%%%%%%%%%%%

%\begin{figure}[ht]
%	\centering
%	\includegraphics[width=80mm]{../plots/cross_section/up_phobic_ggf_bbh1_a1_h1_a1_production.png} 
%	\includegraphics[width=80mm]{../plots/cross_section/up_phobic_ggf_h1_a1_sigmaXbr.png} 
%	\caption{\magenta{(Left) The total production cross-section for heavy Higgs bosons in the gluon-gluon fusion (ggF) channel $(pp \to H_1/A_1, H_1/A_1 \to \tau^+ \tau^-)$ (dashed) together with $b$-associated production $(pp \to b \bar{b} H_1/A_1, H_1/A_1 \to \tau^+ \tau^-)$ (solid), for different benchmark values of $\tan \beta_1$ and $\tan \beta_2$ in the up-phobic limit. The right panel shows the corresponding effective cross section for ggF production $(pp \to H_1/A_1, H_1/A_1 \to \tau^+ \tau^-)$ for several benchmark values of $\tan\beta_1$ and $\tan\beta_2$ in the up-phobic limit. The gray shaded region indicates the current ATLAS limit on the effective cross-section at $13$ TeV with an integrated luminosity of $139$ $fb^{-1}$ ~\cite{ATLAS:2020zms}.}}
%	\label{f:xsec_ggf}
%\end{figure}

%%%%%%%%%%%%%%%%%%%%%%%%%%%%%%%%%%%%%%%%%%%%%%%%%%%%%%%%%%%%
%
\begin{figure}[h!]
	\centering
	\includegraphics[width=80mm,height=60mm]{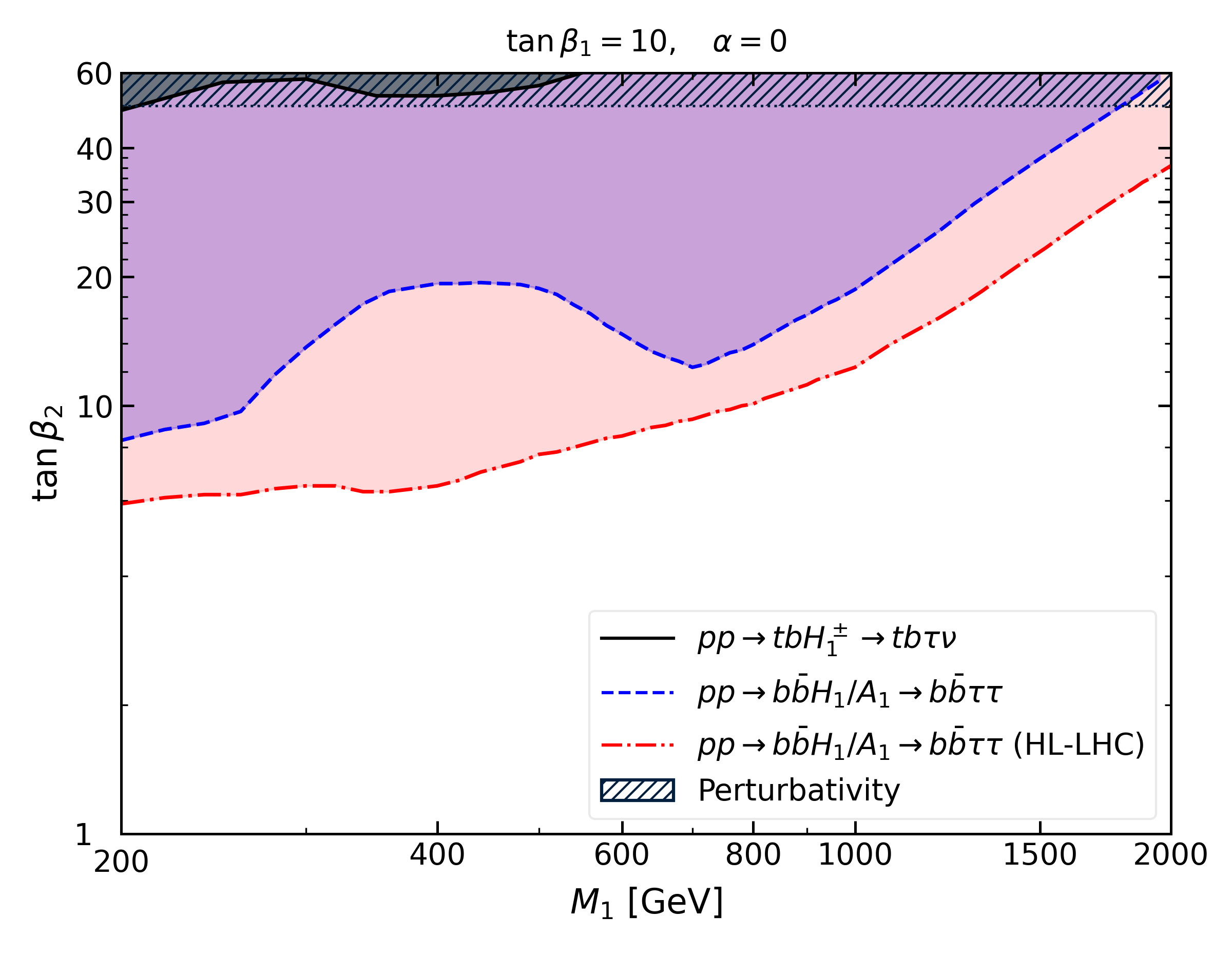} 
	\includegraphics[width=80mm,height=60mm]{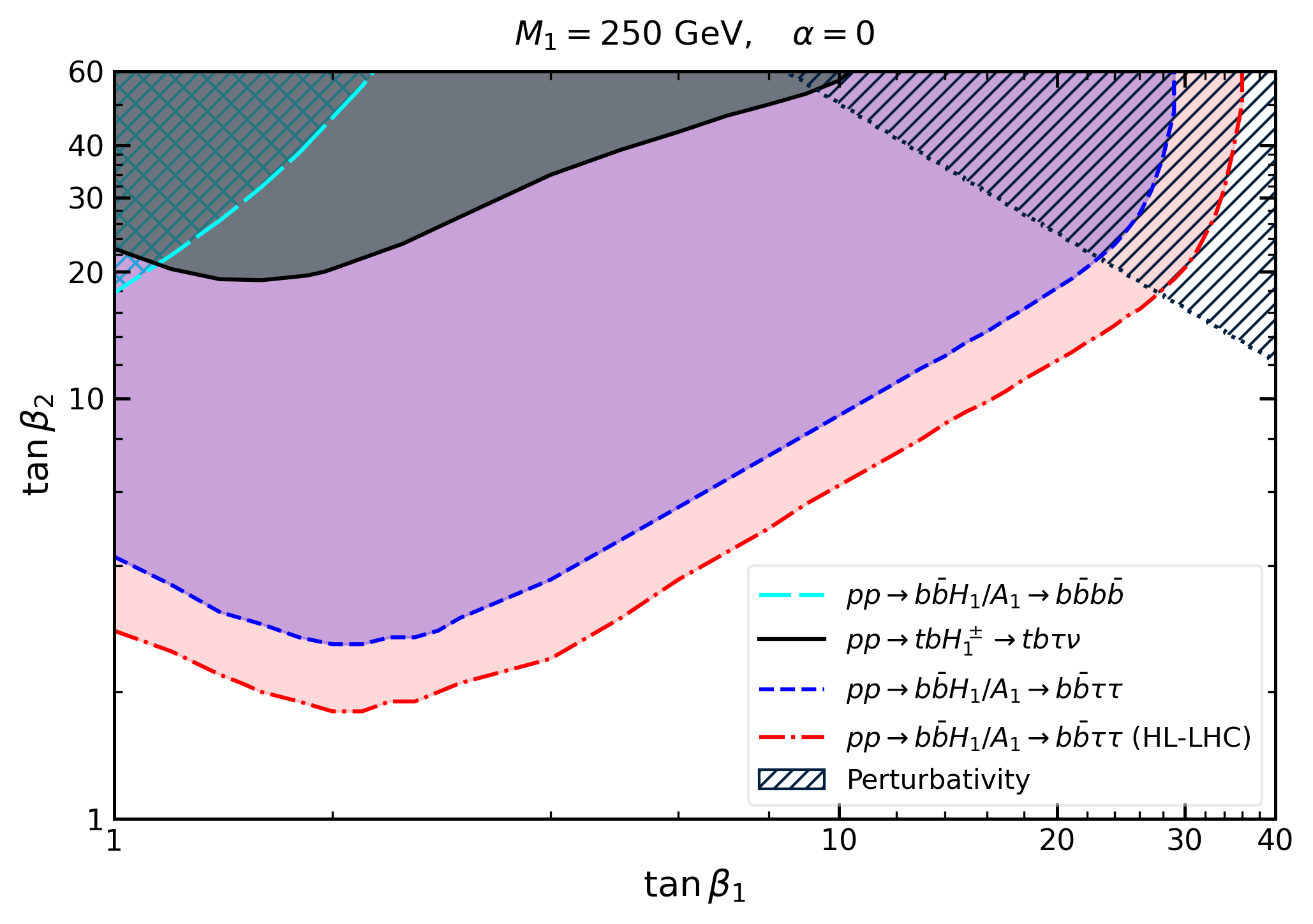} 
	\caption{The combined constraints in the $M_1$--$\tan\beta_2$ plane (left) and in $\tan\beta_1$--$\tan\beta_2$ plane (right) in the up-phobic limit are shown. The shaded black and blue regions denote the excluded parameter space from current ATLAS searches for charged Higgs production, $(pp \to t\bar bH_1^+, H_1^+ \to \tau^+ \nu)$ , and neutral Higgs production, $(pp \to b\bar{b} H_1/A_1, H_1/A_1 \to \tau^+\tau^-)$, respectively. The red shaded region represents the projected constraint from $b$-associated neutral Higgs searches in the $\tau^+\tau^-$ channel at the future HL-LHC~\cite{Cepeda:2019klc}.  The black hatched region in both the figures are excluded from Yukawa perturbativity. }
	\label{f:limit}
\end{figure}
%%%%%%%%%%%%%%%%%%%%%%%%%%%%%%%%%%%%%%%%%%%%%%%%%%%%%%%%%%
%%%%%%%%%   Other Consequences   %%%%%%%%%%%%%%%%%%%%%%%
%%%%%%%%%%%%%%%%%%%%%%%%%%%%%%%%%%%%%%%%%%%%%%%%%%%%%%%%%
%\paragraph{Other consequences:}
%
Finally, to characterize the currently allowed parameter space of the democratic 3HDM in terms of the three relevant parameters, $\tan\beta_{1,2}$ and $M_1$, we present the current LHC constraints in Fig.~\ref{f:limit}. The left panel shows the exclusion limit in the $M_1$--$\tan\beta_2$ plane for a benchmark value of $\tan\beta_1=10$. The current ATLAS search in the $(pp\to b\bar b H_1/A_1$, $H_1/A_1\to \tau^+\tau^-)$ channel already places meaningful constraints on the up-phobic parameter space, excluding a substantial region at large $\tan\beta_2$. The sensitivity is particularly pronounced for relatively light $M_1$, excluding $\tan\beta_2>8$ for $M_1\sim 200$ GeV. To further illustrate the dependence of the search sensitivity on the two $\tan\beta_{1,2}$ parameters, the right panel presents the corresponding constraint in the $\tan\beta_1$--$\tan\beta_2$ plane, fixing the light scalar mass at $M_1=250~\mathrm{GeV}$.  In this case, large $\tan\beta_2$ are disfavored when $\tan\beta_1$ is small, while relatively bigger $\tan\beta_2$ values become allowed when $\tan\beta_1 \gtrsim 10$.
%\sout{The hatched green region is disfavored from the requirement that the total width of the neutral scalars must not exceed its mass, {\it i.e.} $\Gamma(H_1,A_1) < M_1$. }. 
The black hatched region in both the figures are disfavored from the requirement of perturbativity of Yukawa couplings.

For completeness, we also overlay in Fig.~\ref{f:limit} the constraints from the ATLAS search for charged scalars in the $(pp\to t \bar b H_1^+, H_1^+ \to\tau^+ \nu)$ channel~\cite{ATLAS:2024hya}. The corresponding excluded region, shown in black shade, is considerably smaller than that obtained from the neutral-scalar search. 
This comparison highlights the comparatively stronger sensitivity of the $b$-associated neutral-scalar production channel to the parameter space corresponding to the up-phobic scenario.
 We have also shown the existing bounds from the latest CMS result on $pp\to b\bar b H_1/A_1$, $H_1/A_1\to b \bar b$\cite{CMS:2025frm} searches
 as indicated by the cyan shaded region.
%The  existing bounds from the ATLAS searches for $pp\to t \bar bH_1^+, H_1^+\to t \bar b$\cite{ATLAS:2021upq}. This is shown by the \red{color} region.
 However, this search is primarily sensitive to the region with small $\tan\beta_1 \lesssim 1$ and exclude a narrow region in the parameter space.
 To accentuate the effect, we keep the horizontal axis in the right panel in log scale. Additionally, we checked that the existing bounds from the ATLAS searches for $pp\to t \bar bH_1^+, H_1^+\to t \bar b$\cite{ATLAS:2021upq} has no effect in the $\tan\beta_1 > 1$ region and thus is not included in the plot.
 Hence, $b$-associated neutral-scalar production followed by decay to $\tau^+\tau^-$ remains the strongest probe of the up-phobic limit over most of the parameter space.

Furthermore, we show in Fig.~\ref{f:limit}, the projected sensitivity of the HL-LHC in the $pp\to b\bar b H_1/A_1$, $H_1/A_1\to\tau^+\tau^-$ channel~\cite{Cepeda:2019klc}. The future projection suggests a considerable improvement over the current ATLAS sensitivity. The enhanced sensitivity is particularly important for testing the low-mass region and for probing smaller values of $\tan\beta_2$ which are currently inaccessible. 
%%%%%%%%%%%%%%%%%%%%%%%%%%%%%%%%%%%%%%%%%%%%%%%%%%%%%%%%%%%%%%%%%%%%%%%%%%%%%%%%%%%%%%%%%%%%%%%%%%%55

\section{Summary}
\label{sec:summary}
To summarize, we have investigated the up-phobic limit of the democratic Yukawa structure in the 3HDM, focusing on the phenomenology of the nonstandard neutral scalar sector. The up-phobic limit corresponds to the vanishing of the coupling of the nonstandard neutral scalar to up-type quarks. We have derived this limit analytically and shown that it can be realized in the democratic 3HDM through a specific choice of the scalar mixing parameters. Such an up-phobic realization is intrinsically associated with the multi-doublet Yukawa structure of the 3HDM and does not arise as an independent limit in the NFC versions of 2HDM. 
Therefore, this limit can distinguish the democratic 3HDM set-up from its NFC-based 2HDM counterparts.

Motivated by the theoretical constraints from unitarity and BFB, for the LHC phenomenological analysis, we work in the limit of \Eqn{e:theolimit}.
Furthermore, we set $\alpha=0$, to make the up-phobic limit manifest. 
The remaining unknown parameters, namely $M_1$, $M_2$, $\tan\beta_1$ and $\tan\beta_2$ are treated as free.

The simultaneous imposition of the alignment and the up-phobic limits is found to have important consequences for the production of the nonstandard neutral scalars. The couplings of the one-tier of nonstandard neutral scalars to up-type quarks vanish in the up-phobic limit, while alignment suppresses their couplings to the massive gauge bosons. As a result, the conventional production mechanisms that are particularly important for searches for heavy neutral scalars, such as gluon fusion and vector-boson fusion, become strongly suppressed. In contrast, the coupling to bottom quarks can remain sizable and can be enhanced in the relevant region of parameter space. Consequently, $b$-associated production,
$pp\to b\bar b H_1/A_1$, followed by $H_1/A_1\to\tau^+\tau^-$, becomes a particularly important search channel for this scenario. Thus, a channel that is generally regarded as complementary to the conventional neutral-scalar searches acquires a much more prominent role in the up-phobic limit of the democratic 3HDM.

We have also examined the impact of indirect constraints, including those arising from $b\to s\gamma$ and $h\to\gamma\gamma$. These constraints do not provide a strong restriction to the up-phobic parameter space and therefore do not exclude the regions with relatively light nonstandard scalars that remain relevant for direct searches. The direct LHC searches in the $b\bar b H_1/A_1$ production channel with $H_1/A_1\to\tau^+\tau^-$, on the other hand, lead to significant constraints. For a representative benchmark choice of $\tan\beta_1=10$, we find that for $M_1\sim200$ GeV, values of $\tan\beta_2\gtrsim8$ are excluded by the current searches. We further investigate the impact of the LHC searches for a low-mass benchmark of $M_1=250$ GeV in the $\tan\beta_1$ vs $\tan\beta_2$ plane. 
 These results demonstrate that the current $b$-associated $\tau^+\tau^-$ searches already exclude a substantial fraction of the parameter space that would otherwise remain difficult to probe because of the suppression of the conventional production modes.

The phenomenological picture therefore differs substantially from that of standard heavy-neutral-scalar searches. The absence of sizable couplings to up-type quarks and massive vector bosons does not render the nonstandard scalars inaccessible; rather, it shifts the dominant collider sensitivity towards production mechanisms involving bottom quarks.
The future prospects for this channel are equally promising. Improved sensitivity to $b$-associated nonstandard scalar production with $\tau^+\tau^-$ final states can further constrain the currently viable region in the $\tan\beta_1$ vs $\tan\beta_2$ plane identified in this analysis and extend the LHC reach across the up-phobic parameter space.

Our analysis thus demonstrates that, once the conventional production channels are suppressed, the $(pp \to b\bar{b}H_1/A_1,\, H_1/A_1 \to \tau^+\tau^-)$ channel assumes a uniquely important role in probing this otherwise elusive region of parameter space. More broadly, our results establish $b$-associated production as a key experimental handle for testing the up-phobic limit of the democratic 3HDM and for distinguishing this scenario from the NFC versions of 2HDMs.

%%%%%%%%%%%%%%%%%%%%%%%%%%%%%%%%%%%%%%%%
\paragraph{Acknowledgments:}
DD thanks the ANRF (Erstwhile SERB), India for financial support
through grant no. CRG/2022/000565.
IS acknowledges the financial support
through grant no. MTR/2023/00715 from SERB MATRICS project by ANRF (Erstwhile SERB), India.

%%%%%%%%%%%%%%%%%%%%%%%%%%%%%%%%%%%%%%%%%%%%%%%%%%%%%%%%%%%%%%%%%%%%%%%
%%%%%%%%%%%%%%%%%%%%%%%%  Appendix   %%%%%%%%%%%%%%%%%%%%%%%%%%%%%%%%%%
%%%%%%%%%%%%%%%%%%%%%%%%%%%%%%%%%%%%%%%%%%%%%%%%%%%%%%%%%%%%%%%%%%%%%%%%
\section*{Appendix}
\appendix
\setcounter{equation}{0}
\renewcommand{\theequation}{\thesection.\arabic{equation}}
\input{appendix}
\bibliographystyle{JHEP}
\bibliography{bbH}

%\clearpage
%\appendix
%\input{appendix/Z2_Z2_scalarPotential/Z2_Z2_scalarPotential}

\end{document}

%% file: appendix.tex
\section{3HDM Scalar Potential with a softly-broken $Z_2 \times Z_2^\prime$ symmetry}
\label{sec:appendix}

The general form of CP-conserving 3HDM scalar potential with a softy-broken $Z_2\times Z'_2$ symmetry is given as~\cite{Das:2022gbm,Boto:2023nyi}	
\begin{eqnarray}
	V_{Z_2 \times Z'_2} &=& m_{11}^2(\phi_{\bl}^\dagger \phi_{\bl}) +  m_{22}^2(\phi_{\bd}^\dagger 
	\phi_{\bd}) +  m_{33}^2 (\phi_{\bu}^\dagger \phi_{\bu})  - \left(m_{12}^2(\phi_{\bl}^\dagger \phi_{\bd}) +  m_{23}^2(\phi_{\bd}^\dagger 
	\phi_{\bu}) +  m_{13}^2 (\phi_{\bl}^\dagger \phi_{\bu}) + {\rm h.c.} \right)   \nonumber \\
	&&+\lambda_1(\phi_{\bl}^\dagger 
	\phi_{\bl})^2 + \lambda_2(\phi_{\bd}^\dagger \phi_{\bd})^2 + 
	\lambda_3(\phi_{\bu}^\dagger \phi_{\bu})^2  \nonumber \\ 
	&& + \lambda_4 (\phi_{\bl}^\dagger \phi_{\bl})( \phi_{\bd}^\dagger \phi_{\bd}) + \lambda_5 (\phi_{\bl}^\dagger \phi_{\bl})( \phi_{\bu}^\dagger \phi_{\bu}) + 
	\lambda_6 (\phi_{\bd}^\dagger \phi_{\bd})( \phi_{\bu}^\dagger \phi_{\bu}) \nonumber \\
	&& + \lambda_7 (\phi_{\bl}^\dagger \phi_{\bd})( \phi_{\bd}^\dagger \phi_{\bl}) +
	\lambda_8 (\phi_{\bl}^\dagger \phi_{\bu})( \phi_{\bu}^\dagger \phi_{\bl})+ \lambda_9
	(\phi_{\bd}^\dagger \phi_{\bu})( \phi_{\bu}^\dagger \phi_{\bd}) \nonumber \\
	&& + \left[\lambda_{10} (\phi_{\bl}^\dagger \phi_{\bd})^2
	+ \lambda_{11} (\phi_{\bl}^\dagger \phi_{\bu})^2+
	\lambda_{12} (\phi_{\bd}^\dagger \phi_{\bu})^2+ {\rm h.c.} \right] \,,
	\label{e:potential_z2}
\end{eqnarray}
where $\phi_f\,, (f = \bu\,,\bd\,,\bl)$ are the scalar doublets and the field transformation under the $Z_2\times Z_2^\prime$ is given in Eq.~(\ref{e:z2symm}). After the EWSB, the scalar fields can be expanded around their minima as in \Eqn{e:expand_fields} with the parametrization for the VEVs given in \Eqn{e:vev-parametrization}.
%\begin{eqnarray}
%	\phi_f & = & 	\frac{1}{\sqrt{2}}\left(\begin{array}{c}
%		\sqrt{2}	\omega_f^+ \\
%		v_f + h_f + i z_f
%	\end{array}\right),
%\end{eqnarray}
%and the VEVs are parametrized as
%\begin{eqnarray}
%	v_\bu = v \sin \beta_2; \quad v_\bd = v \sin \beta_1 \cos \beta_2; \quad v_\bl = v \cos \beta_1 \cos \beta_2\,. \label{e:vev-parametrization}
%\end{eqnarray}

We assume all parameters of the scalar potential to be real, thereby ensuring CP conservation. Consequently, the neutral scalar states in our case can be classified into CP-even and CP-odd states. This differs from the original Weinberg potential, which accommodates spontaneous CP-violation~\cite{Weinberg:1976hu}. In addition, unlike the Weinberg potential~\cite{Plantey:2022jdg,Ogreid:2024qfw}, we include terms proportional to $m^2_{12}$, $m^2_{23}$, and $m^2_{13}$ that softly break the $Z_2\times Z'_2$ symmetry. These soft-breaking terms are necessary to have non-standard scalars much heavier than the electroweak VEV and thereby ensuring decoupling of such heavy nonstandard scalars ~\cite{Bhattacharyya:2014oka,Faro:2020qyp,Das:2026ldf}.

Using the minimization conditions, the bilinear parameters $m_{11}^2$, $m_{22}^2$, and $m_{33}^2$ can be expressed as:
\begin{subequations}
	\begin{eqnarray}
		m_{11}^2 &=&   -\lambda_1 v_{\bl}^2 -\frac{1}{2}\left\{
		(\lambda_4+\lambda_7 + 2 \lambda_{10})v_{\bd}^2 +(\lambda_5+\lambda_8 + 2 \lambda_{11})v_{\bu}^2 \right\} + m_{12}^2 \frac{v_{\bd}}{v_{\bl}} + m_{13}^2 \frac{v_{\bu}}{v_{\bl}} \,, 
		\\
		m_{22}^2 &=&   -\lambda_2 v_{\bd}^2 -\frac{1}{2}\left\{
		(\lambda_4+\lambda_7 + 2 \lambda_{10})v_{\bl}^2 +(\lambda_6+\lambda_9 + 2 \lambda_{12})v_{\bu}^2 \right\} + m_{12}^2 \frac{v_{\bl}}{v_{\bd}} + m_{23}^2 \frac{v_{\bu}}{v_{\bd}} \,, 
		\\
		m_{33}^2 &=&   -\lambda_3 v_{\bu}^2 -\frac{1}{2}\left\{
		(\lambda_5+\lambda_8 + 2 \lambda_{11})v_{\bl}^2 +(\lambda_6+\lambda_9 + 2 \lambda_{12})v_{\bd}^2 \right\} + m_{13}^2 \frac{v_{\bl}}{v_{\bu}} + m_{23}^2 \frac{v_{\bd}}{v_{\bu}} \,. %\nonumber \\
	\end{eqnarray}
	\label{e:bilinears_z2}
\end{subequations}
We note that the scalar potential in  \Eqn{e:potential_z2} contains 18 independent parameters. The three bilinear parameters $m_{11}^2$, $m_{22}^2$ and $m_{33}^2$ can be traded in favor of the three VEVs, ($v_{\bu\,,\bd\,,\bl}$) as in \Eqn{e:bilinears_z2}. The three VEVs are further parametrized in terms of the total VEV ($v$), $\beta_1$ and $\beta_2$ as in \Eqn{e:vev-parametrization}. The twelve quartic couplings ($\lambda_{1 - 12}$) can then be exchanged in terms of the seven physical scalar masses ($m_h$, $m_{H_{1,2}}$, $m_{A_{1,2}}$ and $m_{H^\pm_{1,2}}$) together with five mixing angles, ($\alpha_{1,2,3}$ and $\gamma_{1,2}$) which we define afterwards. 

Next we introduce three new parameters, $M_1\,, M_2$, and $\alpha$ which will replace the soft-breaking parameters, $m_{12}^2\,, m_{13}^2$ and, $m_{23}^2$ as follows~\cite{Chakraborti:2021bpy}:
\begin{subequations}
	\label{e:soft terms scale}
	\begin{eqnarray}
		m_{12}^2 &=& s_{\beta_1} c_{\beta_1} \left[\left(M^2_1 c^2_{\alpha} + M^2_2 s^2_{\alpha} \right) -s^2_{\beta_2} \left(M^2_1 s^2_{\alpha}+ M^2_{2} c^2_{\alpha}\right)\right] + \left(M_2^2-M_1^2 \right) s_{\beta_2} c_{\alpha} s_{\alpha} c_{2\beta_1} \,, \\
		m_{13}^2 &=& c_{\beta_2}\left[c_{\beta_1} s_{\beta_2}\left(M^2_{2} c^2_{\alpha} + M^2_1 s^2_{\alpha}\right) - \left(M_1^2-M_2^2 \right)  c_{\alpha} s_{\beta_1} s_{\alpha}\right] \,, \\
		m_{23}^2 &=& c_{\beta_2}\left[s_{\beta_1} s_{\beta_2}\left(M^2_{2} c^2_{\alpha} + M^2_1 s^2_{\alpha}\right) + \left(M_1^2-M_2^2 \right)  c_{\beta_1} c_{\alpha} s_{\alpha}\right] \,,
	\end{eqnarray}
\end{subequations}
where, $s_x(c_x)$ is the shorthand for the notation $\sin x(\cos x)$. The relevance of these newly defined parameters will become apparent once we later write the expressions for the self-couplings among the physical scalars.

In the following, we discuss in detail the mass matrices that arise from the scalar potential and the resulting physical mass eigenstates.

\subsection{CP-odd scalar sector} 
Since, we considered a CP-conserving scalar potential, the neutral mass eigenstates can be categorized into CP-even and CP-odd states. 
The  $3\times3$ mass-squared matrix for the pseudoscalar sector (${\cal M}_{P}^2$) can be extracted from the potential as,
\begin{eqnarray}\label{e:cpoddmassmat}
	V_{PS}^{\rm mass} = \begin{pmatrix}
		z_{\bl} & z_{\bd} & z_{\bu}
	\end{pmatrix} \, \frac{{\cal M}_P^2}{2} \, \begin{pmatrix}
		z_{\bl}\\  z_{\bd}\\ z_{\bu}\\
	\end{pmatrix} \,,
\end{eqnarray}
where,
%	\begin{subequations}\label{e:MP2 z2} 
	%		\begin{eqnarray}
		%			{({\cal M}_P^2)}_{11} &=& -2 \left( v_{\bd}^2\lambda_{10}+v_{\bu}^2\lambda_{11}\right) + m_{12}^2\frac{v_{\bd}}{v_{\bl}}+ m_{13}^2\frac{v_{\bu}}{v_{\bl}}\,, \\
		%			{({\cal M}_P^2)}_{12} &=& -m_{12}^2+ 2v_{\bl} v_{\bd}\lambda_{10} \,, \\
		%			{({\cal M}_P^2)}_{13} &=& -m_{13}^2 + 2v_{\bl} v_{\bu} \lambda_{11} \,, \\
		%			{({\cal M}_P^2)}_{22} &=& -2\left(v_{\bl}^2 \lambda_{10}+ v_{\bu}^2 \lambda_{12}\right) + m_{12}^2\frac{v_{\bl}}{v_{\bd}}+ m_{23}^2\frac{v_{\bu}}{v_{\bd}} \,, \\ 
		%			{({\cal M}_P^2)}_{23} &=& -m_{23}^2 + 2 v_{\bd} v_{\bu}\lambda_{12} \,, \\
		%			{({\cal M}_P^2)}_{33} &=&  -2\left(v_{\bl}^2 \lambda_{11}+v_{\bd}^2 \lambda_{12}\right)+ m_{13}^2\frac{v_{\bl}}{v_{\bu}}+ m_{23}^2\frac{v_{\bd}}{v_{\bu}}\,.
		%		\end{eqnarray}
	%	\end{subequations}	
\begin{subequations}\label{e:MP2 z2 scales} 
	\begin{eqnarray}
		{({\cal M}_P^2)}_{11} &=& -2 \left( v_{\bd}^2\lambda_{10}+v_{\bu}^2\lambda_{11}\right)+ \frac{M^2_1\left(v v_{\bd} c_{\alpha}-v_{\bl} v_{\bu} s_{\alpha}\right)^2}{v^2 \left(v_{\bl}^2 +v_{\bd}^2\right)} + \frac{M^2_2 \left(v_{\bl} v_{\bu} c_{\alpha} +v v_{\bd} s_{\alpha}\right)^2}{v^2 \left(v_{\bl}^2 +v_{\bd}^2\right)}\,, \\
		{({\cal M}_P^2)}_{12} &=& 2v_{\bl} v_{\bd}\lambda_{10} -\frac{M^2_1}{v^2\left(v_{\bl}^2 +v_{\bd}^2 \right)}\left[\left(v v_{\bd} c_{\alpha}-v_{\bl} v_{\bu} s_{\alpha}\right) \left(v v_{\bl} c_{\alpha} + v_{\bd} v_{\bu} s_{\alpha}\right)\right] \nonumber  \\ 
		&& +\frac{M^2_2}{v^2\left(v_{\bl}^2 +v_{\bd}^2 \right)}\left[\left(v_{\bd} v_{\bu} c_{\alpha}-v v_{\bl} s_{\alpha}\right) \left(v_{\bl} v_{\bu} c_{\alpha} + v v_{\bd} s_{\alpha}\right)\right]\,, \\
		{({\cal M}_P^2)}_{13} &=&  2v_{\bl} v_{\bu} \lambda_{11} +\frac{M^2_1 s_{\alpha}}{v^2}\left(v v_{\bd} c_{\alpha}-v_{\bl} v_{\bu} s_{\alpha}\right)  -\frac{M^2_2 c_{\alpha}}{v^2}\left(v_{\bl} v_{\bu} c_{\alpha} + v v_{\bd} s_{\alpha}\right)\,, \\
		{({\cal M}_P^2)}_{22} &=& -2\left(v_{\bl}^2 \lambda_{10}+ v_{\bu}^2 \lambda_{12}\right) + \frac{M^2_1\left(v v_{\bl} c_{\alpha}+v_{\bd} v_{\bu} s_{\alpha}\right)^2}{v^2 \left(v_{\bl}^2 +v_{\bd}^2\right)} + \frac{M^2_2 \left(v_{\bd} v_{\bu} c_{\alpha} -v v_{\bl} s_{\alpha}\right)^2}{v^2 \left(v_{\bl}^2 +v_{\bd}^2\right)} \,, \\
		{({\cal M}_P^2)}_{23} &=&  2 v_{\bd} v_{\bu}\lambda_{12} -\frac{M^2_1 s_{\alpha}}{v^2}\left(v v_{\bl} c_{\alpha}+v_{\bd} v_{\bu} s_{\alpha}\right)  +\frac{M^2_2 c_{\alpha}}{v^2}\left(v v_{\bl} s_{\alpha}-v_{\bd} v_{\bu} c_{\alpha}\right) \,, \\
		{({\cal M}_P^2)}_{33} &=&  -2\left(v_{\bl}^2 \lambda_{11}+v_{\bd}^2 \lambda_{12}\right)+\frac{\left(v_{\bl}^2 +v_{\bd}^2\right)}{v^2}\left(M^2_1 s^2_{\alpha} +M^2_2 c^2_{\alpha}\right)\,.
	\end{eqnarray}
\end{subequations}	
This mass-squared matrix can be block diagonalized as follows:
%\footnote{Such a block diagonalization
	%   in the pseudoscalar and the charged scalar sector is a general
	%   property of CP conserving 3HDMs.}

\begin{eqnarray}\label{e:mp2x2 }
	{\cal B}_{P}^2 \equiv {\cal O}_{\beta} \cdot {\cal M}_{P}^2 \cdot {\cal O}_{\beta}^{T} &=& \begin{pmatrix}
		0 & 0 & 0 \\
		0 & {({\cal B}_P^2)}_{22}  & {({\cal B}_P^2)}_{23} \\
		0 & {({\cal B}_P^2)}_{23} & {({\cal B}_P^2)}_{33} \\
	\end{pmatrix} \,,
\end{eqnarray}
\begin{subequations}\label{e:BP2 z2 scales}
	where $\mathcal{O}_\beta$ is defined in Eq.~(\ref{e:obeta}). The non-vanishing elements of ${{\cal B}_P^2}$ are given by,
	\begin{eqnarray}
		{({\cal B}_P^2)}_{22} &=& -\frac{2}{\left(v_{\bl}^2+v_{\bd}^2\right)}  \left[ \lambda_{10} \left(v_{\bl}^2+v_{\bd}^2\right)^2+ \left(\lambda_{11} v_{\bd}^2  + \lambda_{12} v_{\bl}^2 \right)v_{\bu}^2 \right]+M^2_1 c^2_{\alpha} + M^2_ 2 s^2_{\alpha}\,, 
		\\
		{({\cal B}_P^2)}_{23}  &=&\frac{-2v v_{\bl} v_{\bd} v_{\bu}}{\left(v_{\bl}^2+v_{\bd}^2\right)} \left( \lambda_{11}-\lambda_{12}\right) - \frac{\left( M^2_1 -M^2_2 \right)s_{2 \alpha} }{2}  \,,\\ 
		{({\cal B}_P^2)}_{33} &=&  -\frac{2 v^2}{\left(v_{\bl}^2+v_{\bd}^2\right)} \left(\lambda_{11} v_{\bl}^2 + \lambda_{12} v_{\bd}^2 \right)+  M^2_1 s^2_{\alpha} +M^2_2 c^2_{\alpha}  \,.
	\end{eqnarray}
\end{subequations}
The remaining $2\times2$ block of ${{\cal B}_P^2}$ can be assumed to be fully diagonalized by an orthogonal transformation as (see \Eqn{e:pseudorot})
%\begin{subequations}
\begin{eqnarray}
	\begin{pmatrix}		
		\cos\gamma_1 & -\sin\gamma_1 \\
		\sin\gamma_1 & \cos\gamma_1 \end{pmatrix} \cdot \begin{pmatrix}
		{({\cal B}_P^2)}_{22}  & {({\cal B}_P^2)}_{23} \\
		{({\cal B}_P^2)}_{23} & {({\cal B}_P^2)}_{33} \\
	\end{pmatrix} \cdot \begin{pmatrix}
		\cos\gamma_1 & \sin\gamma_1 \\
		-\sin\gamma_1 & \cos\gamma_1 \end{pmatrix} &=& 
	\iffalse
	\begin{pmatrix}
		0 & 0 & 0 \\
		0 & m_{A_1}^2 & 0 \\
		0 & 0 & m_{A_2}^2 \\
	\end{pmatrix}
	\fi
	{\rm diag}  (~ m^2_{A_1},~ m^2_{A_2})
	\,,\label{e:PSrot} 
\end{eqnarray}
%where, ${\cal O}_{\gamma_1}$ is given in Eq.~(\ref{eq:gammaRots}), 
%	\begin{eqnarray}
	%{\cal O}_{\gamma_1} = 
	%\begin{pmatrix}
	%1 & 0 & 0 \\ 
	%0 & \cos\gamma_1 & -\sin\gamma_1 \\ 
	%0 & \sin\gamma_1 & \cos\gamma_1 \end{pmatrix}  \label{e:Ogamma1} \,.
%\end{eqnarray}
%\end{subequations}
which entails the following relations
\begin{subequations}\label{e:mAtoBP2}
	\begin{eqnarray}
		m^2_{A_1} \cos^2 \gamma_1 +  m^2_{A_2} \sin^2 \gamma_1 &=& {({\cal B}_P^2)}_{22}   \,, \\
		\cos \gamma_1 \sin\gamma_1 (m^2_{A_2} - m^2_{A_1})  &=&	{({\cal B}_P^2)}_{23} \,, \\
		m^2_{A_1} \sin^2 \gamma_1 +  m^2_{A_2} \cos^2 \gamma_1 &=& {({\cal B}_P^2)}_{33}  \,.
	\end{eqnarray}
\end{subequations}
Using Eq.~(\ref{e:BP2 z2 scales}) and Eq.~(\ref{e:mAtoBP2}) we can solve for $\lambda_{10}\,, \lambda_{11}$ and $\lambda_{12}$ as 
\begin{subequations}\label{e:lam12 z2 scales}
	\begin{eqnarray}
		\lambda_{10} & =& 
		\frac{1}{2 v^2}
		\Bigg[ \frac{1}{c^2_{\beta_2}}
		\left\{
		M_1^2 c^2_{\alpha}
		+ M_2^2 s^2_{\alpha}
		-
		\left(
		m_{A_1}^2 c^2_{\gamma_1}
		+ m_{A_2}^2 s^2_{\gamma_1}
		\right)
		\right\} \nonumber \\
		&&
		+ \frac{\tan\beta_2}{\tan{2 \beta_1} c_{\beta_2}}
		\left\{
		\left(m_{A_1}^2 - m_{A_2}^2\right)s_{2 \gamma_1}
		-
		\left(M_1^2 - M_2^2\right)s_{2 \alpha}
		\right\}
		\nonumber\\
		&&
		+ \tan^2\beta_2
		\left\{
		m_{A_1}^2 s^2_{\gamma_1} +m_{A_2}^2 c^2_{\gamma_1} -\left(M_1^2 s^2_{\alpha} + M_2^2 c^2_{\alpha} \right)
		\right\}
		\Bigg]\,,\\
		\lambda_{11} &=&
		\frac{1}{2 v^2}
		\Bigg[
		\left\{
		M_1^2 s^2_{\alpha}
		+ M_2^2 c^2_{\alpha}
		-
		\left(
		m_{A_1}^2 s^2_{\gamma_1}
		+ m_{A_2}^2 c^2_{\gamma_1}
		\right)
		\right\} \nonumber \\
		&&
		+ \frac{\tan\beta_1}{2 s_{\beta_2}}
		\left\{
		\left(m_{A_1}^2 - m_{A_2}^2\right)s_{2 \gamma_1}
		-
		\left(M_1^2 - M_2^2\right)s_{2 \alpha}
		\right\}
		\Bigg]\,,\\
		\lambda_{12} &=&
		\frac{1}{2 v^2}
		\Bigg[
		\left\{
		M_1^2 s^2_{\alpha}
		+ M_2^2 c^2_{\alpha}
		-
		\left(
		m_{A_1}^2 s^2_{\gamma_1}
		+ m_{A_2}^2 c^2_{\gamma_1}
		\right)
		\right\} \nonumber \\
		&&
		- \frac{\cot\beta_1}{2 s_{\beta_2}}
		\left\{
		\left(m_{A_1}^2 - m_{A_2}^2\right)s_{2 \gamma_1}
		-
		\left(M_1^2 - M_2^2\right)s_{2 \alpha}
		\right\}
		\Bigg]\,.
	\end{eqnarray}
\end{subequations}

%%%%%%%%%%%%%%%%%%%%%%%%%%%%%%%%%%%%%%%%%%%%%%%%%%%%%%%%%%%%%%%%%
%%%%%%%%%           charged scalar sector         %%%%%%%%%%%%%%
%%%%%%%%%%%%%%%%%%%%%%%%%%%%%%%%%%%%%%%%%%%%%%%%%%%%%%%%%%%%%%%%

\subsection{Charged scalar sector} 
The mass term for the  charged scalar sector can be extracted from the scalar potential as,
\begin{eqnarray}\label{e:chargedmassmat}
	V_{C}^{\rm mass} = \begin{pmatrix}
		w_{\bl}^- & w_{\bd}^- & w_{\bu}^-
	\end{pmatrix} \,{\cal M}_C^2 \, \begin{pmatrix}
		w_{\bl}^+\\  w_{\bd}^+\\ w_{\bu}^+\\
	\end{pmatrix} \, + {\rm h.c.}.
\end{eqnarray}
Similar to the pseudoscalar case, the elements of $3\times3$ charged sector mass matrix ${\cal M}_{C}^2$ are given as follows
\begin{subequations}\label{e:MC2 z2}
	\begin{eqnarray}
	{({\cal M}_C^2)}_{11} &=& - \frac{1}{2}\left[ v_{\bd}^2\left (\lambda_7  +2 \lambda_{10}\right)+ v_{\bu}^2 \left(\lambda_8   +2 \lambda_{11}\right )\right]+ \frac{M^2_1\left(v v_{\bd} c_{\alpha}-v_{\bl} v_{\bu} s_{\alpha}\right)^2}{v^2 \left(v_{\bl}^2 +v_{\bd}^2\right)} \nonumber \\
	&&  + \frac{M^2_2 \left(v_{\bl} v_{\bu} c_{\alpha} +v v_{\bd} s_{\alpha}\right)^2}{v^2 \left(v_{\bl}^2 +v_{\bd}^2\right)}, \\
	{({\cal M}_C^2)}_{12} &=& \frac{v_{\bl} v_{\bd}}{2}\left(\lambda_7 + 2\lambda_{10}\right) -\frac{M^2_1}{v^2\left(v_{\bl}^2 +v_{\bd}^2 \right)}\left[\left(v v_{\bd} c_{\alpha}-v_{\bl} v_{\bu} s_{\alpha}\right) \left(v v_{\bl} c_{\alpha} + v_{\bd} v_{\bu} s_{\alpha}\right)\right] \nonumber  \\ 
	&&  
	+\frac{M^2_2}{v^2\left(v_{\bl}^2 +v_{\bd}^2 \right)}\left[\left(v_{\bd} v_{\bu} c_{\alpha}-v v_{\bl} s_{\alpha}\right) \left(v_{\bl} v_{\bu} c_{\alpha} + v v_{\bd} s_{\alpha}\right)\right]\,, \\
	{({\cal M}_C^2)}_{13} &=& \frac{v_{\bl} v_{\bu}}{2}\left(\lambda_8 +2\lambda_{11}\right) +\frac{M^2_1 s_{\alpha}}{v^2}\left(v v_{\bd} c_{\alpha}-v_{\bl} v_{\bu} s_{\alpha}\right) 
	-\frac{M^2_2 c_{\alpha}}{v^2}\left(v_{\bl} v_{\bu} c_{\alpha} + v v_{\bd} s_{\alpha}\right)  \,, \\
	{({\cal M}_C^2)}_{22} &=&   - \frac{1}{2}\left[ v_{\bl}^2\left (\lambda_7  +2 \lambda_{10}\right)+ v_{\bu}^2 \left(\lambda_9   +2 \lambda_{12}\right )\right]+  \frac{M^2_1\left(v v_{\bl} c_{\alpha}+v_{\bd} v_{\bu} s_{\alpha}\right)^2}{v^2 \left(v_{\bl}^2 +v_{\bd}^2\right)} \nonumber \\
	&& + \frac{M^2_2 \left(v_{\bd} v_{\bu} c_{\alpha} -v v_{\bl} s_{\alpha}\right)^2}{v^2 \left(v_{\bl}^2 +v_{\bd}^2\right)} \,, \\ 
	{({\cal M}_C^2)}_{23} &=& \frac{v_{\bd} v_{\bu}}{2}\left(\lambda_9 +2 \lambda_{12}\right)-\frac{M^2_1 s_{\alpha}}{v^2}\left(v v_{\bl} c_{\alpha}+v_{\bd} v_{\bu} s_{\alpha}\right) +\frac{M^2_2 c_{\alpha}}{v^2}\left(  v v_{\bl} s_{\alpha}-v_{\bd} v_{\bu} c_{\alpha}\right) \,, \\
	{({\cal M}_C^2)}_{33} &=&  - \frac{1}{2}\left[ v_{\bl}^2\left (\lambda_8  +2 \lambda_{11}\right)+ v_{\bd}^2 \left(\lambda_9  +2 \lambda_{12}\right )\right]+\frac{\left(v_{\bl}^2 +v_{\bd}^2\right)}{v^2}\left(M^2_1 s^2_{\alpha} +M^2_2 c^2_{\alpha}\right) \,.
\end{eqnarray}
\end{subequations}
This ${\cal M}_{C}^2$ can be block diagonalized using the $\mathcal{O}_\beta$ matrix of Eq.~(\ref{e:obeta}) as follows:
% --------------------------------------------------
\begin{subequations}\label{e:BC2 z2}
	\begin{eqnarray}
		{\cal B}_{C}^2 \equiv {\cal O}_{\beta} \cdot {\cal M}_{C}^2 \cdot {\cal O}_{\beta}^{T} &=& \begin{pmatrix}
			0 & 0 & 0 \\
			0 & {({\cal B}_C^2)}_{22}  & {({\cal B}_C^2)}_{23} \\
			0 & {({\cal B}_C^2)}_{23} & {({\cal B}_C^2)}_{33} \\
		\end{pmatrix} \,,
	\end{eqnarray}
	where,
	%	\begin{eqnarray}
		%		{({\cal B}_C^2)}_{22} &=& -\frac{1}{2(v_{\bl}^2+v_{\bd}^2)} \bigg[  \left(\lambda_7 + 2\lambda_{10}\right )\left(v_{\bl}^2+v_{\bd}^2\right)^2+v_{\bd}^2 v_{\bu}^2 \left(\lambda_8 +2 \lambda_{11} \right) +v_{\bl}^2  v_{\bu}^2 \left(\lambda_9 +2\lambda_{12} \right)\nonumber \\
		%		&&  - 2 m_{12}^2\frac{(v_{\bl}^2 + v_{\bd}^2)^2}{v_{\bl} v_{\bd}} - 2 m_{13}^2\frac{ v_{\bd}^2 v_{\bu} }{v_{\bl}} - 2 m_{23}^2\frac{ v_{\bl}^2 v_{\bu} }{v_{\bd}}  \bigg]\,. \\
		%		{({\cal B}_C^2)}_{23}  &=& \frac{v}{2(v_{\bl}^2+v_{\bd}^2)} \left[ v_{\bl} v_{\bd} v_{\bu} \left (-\lambda_8 +\lambda_9 -2 \lambda_{11} +2\lambda_{12}\right) - 2 m_{23}^2 v_{\bl} + 2 m_{13}^2 v_{\bd} \right] \,, \\
		%		{({\cal B}_C^2)}_{33} &=& -\frac{v^2}{2(v_{\bl}^2+v_{\bd}^2)} \left[v_{\bl}^2(\lambda_8 +2\lambda_{11}) +v_{\bd}^2(\lambda_9 + 2\lambda_{12}) - 2 m_{13}^2 \frac{v_{\bl}}{v_{\bu}}  - 2 m_{23}^2 \frac{v_{\bd}}{v_{\bu}} \right] \,,
		%	\end{eqnarray}
	\begin{eqnarray}
		{({\cal B}_C^2)}_{22} &=& -\frac{1}{2(v_{\bl}^2+v_{\bd}^2)} \left[  \left(\lambda_7 + 2\lambda_{10}\right )\left(v_{\bl}^2+v_{\bd}^2\right)^2+v_{\bd}^2 v_{\bu}^2 \left(\lambda_8 +2 \lambda_{11} \right) +v_{\bl}^2  v_{\bu}^2 \left(\lambda_9 +2\lambda_{12} \right)  \right]\nonumber \\
		&& 
		+ M^2_1 c^2_{\alpha} + M^2_2 s^2_{\alpha}\,. \\
		{({\cal B}_C^2)}_{23}  &=& \frac{v v_{\bl} v_{\bd} v_{\bu}}{2(v_{\bl}^2+v_{\bd}^2)} \left[ -\lambda_8 +\lambda_9 -2 \lambda_{11} +2\lambda_{12} \right]- \frac{\left( M^2_1 -M^2_2 \right)s_{2 \alpha} }{2} \,, \\
		{({\cal B}_C^2)}_{33} &=& -\frac{v^2}{2(v_{\bl}^2+v_{\bd}^2)} \left[v_{\bl}^2(\lambda_8 +2\lambda_{11}) +v_{\bd}^2(\lambda_9 + 2\lambda_{12}) \right] +  M^2_1 s^2_{\alpha} +M^2_2 c^2_{\alpha}\,.
	\end{eqnarray}
\end{subequations}
%-----------------------------------------------------------
Further, the remaining $2\times2$ block can be completely diagonalized  with the rotation (see \Eqn{e:chargedrot})
\iffalse
\begin{subequations}\label{e:BCrot}
	\begin{eqnarray}
		{\cal O}_{\gamma_2} \cdot ({\cal B}_{C})^2 \cdot {\cal O}_{\gamma_2}^T = \left(\begin{array}{ccc}
			0 & 0 & 0 \\
			0 & m_{H_1^+}^2 & 0 \\
			0 & 0 & m_{H_2^+}^2 \\
		\end{array}\right)\,,
	\end{eqnarray}
	%	where,
	%	\begin{eqnarray}
		%{\cal O}_{\gamma_2} = 
		%\begin{pmatrix}
		%1 & 0 & 0 \\ 
		%0 & \cos\gamma_2 & -\sin\gamma_2 \\ 
		%0 & \sin\gamma_2 & \cos\gamma_2 \end{pmatrix}  \label{e:Ogamma2} \,.
	%\end{eqnarray}
\end{subequations}
\fi
%\begin{subequations}
\begin{eqnarray}
	\begin{pmatrix}		
		\cos\gamma_2 & -\sin\gamma_2 \\
		\sin\gamma_2 & \cos\gamma_2 \end{pmatrix} \cdot \begin{pmatrix}
		{({\cal B}_C^2)}_{22}  & {({\cal B}_C^2)}_{23} \\
		{({\cal B}_C^2)}_{23} & {({\cal B}_C^2)}_{33} \\
	\end{pmatrix} \cdot \begin{pmatrix}
		\cos\gamma_2 & \sin\gamma_2 \\
		-\sin\gamma_2 & \cos\gamma_2 \end{pmatrix} &=& 
	\iffalse
	\begin{pmatrix}
		m_{H^{\pm}_1}^2 & 0 \\
		0 & m_{H^{\pm}_2}^2 \\
	\end{pmatrix}
	\fi
	{\rm diag}(~ m^2_{H^{\pm}_1},~ m^2_{H^{\pm}_2})
	\,.\label{e:PSrot} 
\end{eqnarray}
Thus, we will have the following relations:
\begin{subequations}\label{e:mctoapbpcp}
	\begin{eqnarray}
		m^2_{H_1^{\pm}} \cos^2 \gamma_2 +  m^2_{H_2^{\pm}} \sin^2 \gamma_2 &=& {({\cal B}_C^2)}_{22}\,, \\
		\cos \gamma_2 \sin\gamma_2 (m^2_{H_2^{\pm}} - m^2_{H_1^{\pm}}) &=&  {({\cal B}_C^2)}_{23}\,, \\
		m^2_{H_1^{\pm}} \sin^2 \gamma_2 +  m^2_{H_2^{\pm}} \cos^2 \gamma_2 &=&  {({\cal B}_C^2)}_{33}\,.
	\end{eqnarray}
\end{subequations}
These equations in conjunction with Eq.~(\ref{e:BC2 z2}) enable us to solve for $\lambda_{7},\lambda_8$, and $\lambda_9$ as given below:
\begin{subequations} \label{e:lam789 z2}
	%	\begin{eqnarray}
		%		\lambda_7  &=& %\frac{m_{C1}^2}{2 v^2}  \left[-2 + (-3 + c_{2\beta_2})\frac{c_{2\gamma_2}}{c_{\beta_2}^2} + \frac{4\tan \beta_2}{\tan 2\beta_1} \frac{ s_{2\gamma_2}}{c_{\beta_2}}\right]
		%		%-\frac{m_{C2}^2}{2 v^2} \left[2 +(-3 + c_{2\beta_2})\frac{c_{2\gamma_2}}{c_{\beta_2}^2} + \frac{4\tan \beta_2}{\tan 2\beta_1} \frac{s_{2\gamma_2}}{c_{\beta_2}}\right]  \nonumber \\
		%		\frac{\left(m_{H_1^+}^2 - m_{H_2^+}^2\right)}{2 v^2} \left[(-3 + c_{2\beta_2})\frac{c_{2\gamma_2}}{c_{\beta_2}^2} + \frac{4\tan \beta_2}{\tan 2\beta_1} \frac{ s_{2\gamma_2}}{c_{\beta_2}}\right] -
		%		\frac{\left(m_{H_1^+}^2 + m_{H_2^+}^2\right)}{v^2} \nonumber \\
		%		&& - 2 \lambda_{10}  +  \frac{2 m_{12}^2 }{ v^2 s_{\beta_1} c_{\beta_1} c_{\beta_2}^2 }\,, \\
		%		%
		%		\lambda_8  &=&\frac{m_{H_1^+}^2}{v^2}\left(-2s^2_{\gamma_2}+\tan\beta_1 \frac{s_{2\gamma_2} }{s_{\beta_2} }\right)
		%		-\frac{m_{H_2^+}^2}{v^2}\left(2c^2_{\gamma_2}+\tan\beta_1  \frac{ s_{2\gamma_2}}{s_{\beta_2}}\right)  - 2\lambda_{11} \nonumber \\
		%		&&
		%		+  \frac{2 m_{13}^2 }{ v^2 c_{\beta_1} s_{\beta_2} c_{\beta_2}}\,,  \\
		%		%
		%		\lambda_9  &=& -\frac{m_{H_1^+}^2}{v^2}\left(2s^2_{\gamma_2}+\cot\beta_1 \frac{s_{2\gamma_2} }{ s_{\beta_2}}\right)
		%		+\frac{m_{H_2^+}^2}{v^2}\left(-2c^2_{\gamma_2}+\cot\beta_1  \frac{s_{2\gamma_2} }{s_{\beta_2}}\right) -2\lambda_{12}\nonumber \\
		%		&&
		%		+  \frac{2 m_{23}^2 }{ v^2 s_{\beta_1} s_{\beta_2} c_{\beta_2}}\,. 
		%	\end{eqnarray}
	\begin{eqnarray}
		\lambda_7  &=& \frac{1}{v^2}\Big[\left\{\left(M_1^2 +M_2^2\right)-\left( m_{H_1^{\pm}}^2 + m_{H_2^{\pm}}^2\right) \right\} \nonumber \\
		&& +
		\frac{\left(-3 + c_{2\beta_2}\right)}{2 c_{\beta_2}^2}\left\{\left(m_{H_1^{\pm}}^2 -m_{H_2^{\pm}}^2\right)c_{2\gamma_2} - \left(M_1^2 -M_2^2\right)c_{2\alpha}\right\}\nonumber \\
		&&
		+\frac{2 \tan\beta_2}{\tan2\beta_1 c_{\beta_2}}\left\{\left(m_{H_1^{\pm}}^2 -m_{H_2^{\pm}}^2\right)s_{2\gamma_2} - \left(M_1^2 -M_2^2\right)s_{2\alpha}\right\} \Big] - 2 \lambda_{10}  
		\,, \\
		\lambda_8  &=&
		\frac{2}{ v^2}
		\Bigg[
		\left\{
		M_1^2 s^2_{\alpha}
		+ M_2^2 c^2_{\alpha}
		-
		\left(
		m_{H_1^{\pm}}^2 s^2_{\gamma_2}
		+ m_{H_2^{\pm}}^2 c^2_{\gamma_2}
		\right)
		\right\} \nonumber \\
		&&
		+ \frac{\tan\beta_1}{2 s_{\beta_2}}
		\left\{
		\left(m_{H_1^{\pm}}^2 - m_{H_2^{\pm}}^2\right)s_{2 \gamma_2}
		-
		\left(M_1^2 - M_2^2\right)s_{2 \alpha}
		\right\}
		\Bigg] - 2\lambda_{11} \,,  \\
		\lambda_9  &=& 
		\frac{2}{ v^2}
		\Bigg[
		\left\{
		M_1^2 s^2_{\alpha}
		+ M_2^2 c^2_{\alpha}
		-
		\left(
		m_{H_1^{\pm}}^2 s^2_{\gamma_2}
		+ m_{H_2^{\pm}}^2 c^2_{\gamma_2}
		\right)
		\right\} \nonumber \\
		&&
		+ \frac{\cot\beta_1}{2 s_{\beta_2}}
		\left\{
		\left(M_1^2 - M_2^2\right)s_{2 \alpha} - \left(m_{H_1^{\pm}}^2 - m_{H_2^{\pm}}^2\right)s_{2 \gamma_2}
		\right\}
		\Bigg] - 2\lambda_{12}\,. 
	\end{eqnarray}
\end{subequations}
where, the other three couplings $\lambda_{10}, \lambda_{11 }\, \text{and} ~\lambda_{12}$ can be replaced using Eq.~(\ref{e:lam12 z2 scales}). 

%%%%%%%%%%%%%%%%%%%%%%%%%%%%%%%%%%%%%%%%%%%%%%%%%%%
%%%%%%%%%%  CP even sector  %%%%%%%%%%%%%%%%%%%
%%%%%%%%%%%%%%%%%%%%%%%%%%%%%%%%%%%%%%%%%%%%%%%%

\subsection{CP-even scalar sector}
The mass terms in the neutral scalar sector can be extracted from the potential as,
\begin{eqnarray}\label{e:neutralmassmat}
	V_{S}^{\rm mass} =\begin{pmatrix}
		h_{\bl} & h_{\bd} & h_{\bu}\\
	\end{pmatrix} \frac{{\cal M}_S^2}{2} \begin{pmatrix}
		h_{\bl}\\  h_{\bd}\\ h_{\bu}\\
	\end{pmatrix} \,,
\end{eqnarray}
where, ${\cal M}_S^2$ is the $3\times3$ symmetric
mass matrix whose elements are given by,
\begin{subequations} 
	%		\begin{eqnarray}
		%			{({\cal M}_S^2)}_{11} &=& 2 v_{\bl}^2 \lambda_1 + m_{12}^2\frac{v_{\bd}}{v_{\bl}} + m_{13}^2\frac{v_{\bu}}{v_{\bl}} \,, \\
		%			{({\cal M}_S^2)}_{12} &=&  v_{\bl} v_{\bd}\left (\lambda_4+\lambda_{7}+ 2\lambda_{10}\right) - m_{12}^2 \,, \\
		%			{({\cal M}_S^2)}_{13} &=&  v_{\bl} v_{\bu}\left (\lambda_5+\lambda_{8}+ 2\lambda_{11}\right) - m_{13}^2  \,, \\
		%			{({\cal M}_S^2)}_{22} &=& 2 v_{\bd}^2 \lambda_2  + m_{12}^2\frac{v_{\bl}}{v_{\bd}} + m_{23}^2\frac{v_{\bu}}{v_{\bd}}\,, \\ 
		%			{({\cal M}_S^2)}_{23} &=& v_{\bd} v_{\bu}\left (\lambda_6+\lambda_{9}+ 2\lambda_{12}\right) - m_{23}^2 \,, \\
		%			{({\cal M}_S^2)}_{33} &=& 2 v_{\bu}^2 \lambda_3 + m_{13}^2\frac{v_{\bl}}{v_{\bu}} + m_{23}^2\frac{v_{\bd}}{v_{\bu}} \,.
		%		\end{eqnarray}\label{e:ms element z2}
	\begin{eqnarray}
		{({\cal M}_S^2)}_{11} &=& 2 v_{\bl}^2 \lambda_1 + \frac{M^2_1\left(v v_{\bd} c_{\alpha}-v_{\bl} v_{\bu} s_{\alpha}\right)^2}{v^2 \left(v_{\bl}^2 +v_{\bd}^2\right)} + \frac{M^2_2 \left(v_{\bl} v_{\bu} c_{\alpha} +v v_{\bd} s_{\alpha}\right)^2}{v^2 \left(v_{\bl}^2 +v_{\bd}^2\right)}  \,, \\
		{({\cal M}_S^2)}_{12} &=&  v_{\bl} v_{\bd}\left (\lambda_4+\lambda_{7}+ 2\lambda_{10}\right) -\frac{M^2_1}{v^2\left(v_{\bl}^2 +v_{\bd}^2 \right)}\left[\left(v v_{\bd} c_{\alpha}-v_{\bl} v_{\bu} s_{\alpha}\right) \left(v v_{\bl} c_{\alpha} + v_{\bd} v_{\bu} s_{\alpha}\right)\right] \nonumber  \\ 
		&& +\frac{M^2_2}{v^2\left(v_{\bl}^2 +v_{\bd}^2 \right)}\left[\left(v_{\bd} v_{\bu} c_{\alpha}-v v_{\bl} s_{\alpha}\right) \left(v_{\bl} v_{\bu} c_{\alpha} + v v_{\bd} s_{\alpha}\right)\right] \,, \\
		{({\cal M}_S^2)}_{13} &=&  v_{\bl} v_{\bu}\left (\lambda_5+\lambda_{8}+ 2\lambda_{11}\right) +\frac{M^2_1 s_{\alpha}}{v^2}\left(v v_{\bd} c_{\alpha}-v_{\bl} v_{\bu} s_{\alpha}\right) \nonumber \\
		&& -\frac{M^2_2 c_{\alpha}}{v^2}\left(v_{\bl} v_{\bu} c_{\alpha} + v v_{\bd} s_{\alpha}\right) \,, \\
		{({\cal M}_S^2)}_{22} &=& 2 v_{\bd}^2 \lambda_2  + \frac{M^2_1\left(v v_{\bl} c_{\alpha}+v_{\bd} v_{\bu} s_{\alpha}\right)^2}{v^2 \left(v_{\bl}^2 +v_{\bd}^2\right)} + \frac{M^2_2 \left(v_{\bd} v_{\bu} c_{\alpha} -v v_{\bl} s_{\alpha}\right)^2}{v^2 \left(v_{\bl}^2 +v_{\bd}^2\right)}\,, \\ 
		{({\cal M}_S^2)}_{23} &=& v_{\bd} v_{\bu}\left (\lambda_6+\lambda_{9}+ 2\lambda_{12}\right) -\frac{M^2_1 s_{\alpha}}{v^2}\left(v v_{\bl} c_{\alpha}+v_{\bd} v_{\bu} s_{\alpha}\right)\nonumber \\
		&& +\frac{M^2_2 c_{\alpha}}{v^2}\left( v v_{\bl} s_{\alpha}-v_{\bd} v_{\bu} c_{\alpha}\right) \,, \\
		{({\cal M}_S^2)}_{33} &=& 2 v_{\bu}^2 \lambda_3 +\frac{\left(v_{\bl}^2 +v_{\bd}^2\right)}{v^2}\left(M^2_1 s^2_{\alpha} +M^2_2 c^2_{\alpha}\right) \,.
	\end{eqnarray}\label{e:mselement z2 scales}
\end{subequations}	
This mass matrix should be diagonalized via the following orthogonal transformation
\begin{eqnarray}\label{e:msdiag z2}
	{\cal O}_\alpha \cdot {\cal M}_{S}^2 \cdot {\cal O}_\alpha^T = {\rm diag} (m_h^2,~m_{H_1}^2,~ m_{H_2}^2)\,,
\end{eqnarray}
where, ${\cal O}_\alpha$ is defined as (compare with \Eqn{e:cpevenrot})
\begin{eqnarray}\label{e:Oa}
	{\cal O}_\alpha &=& \begin{pmatrix}
		1 & 0 & 0 \\
		0 & \cos \alpha_3 &  -\sin \alpha_3  \\
		0 & \sin \alpha_3 & \cos \alpha_3 
	\end{pmatrix}
	\begin{pmatrix}
		\cos \alpha_2 & 0 & \sin \alpha_2  \\
		0 & 1 & 0 \\
		-\sin \alpha_2 & 0 & \cos \alpha_2 
	\end{pmatrix} \begin{pmatrix}
		\cos \alpha_1 & \sin \alpha_1 & 0 \\
		-\sin \alpha_1 & \cos \alpha_1 & 0 \\
		0 & 0 & 1 \end{pmatrix} \,.
\end{eqnarray}
Inverting Eq.~(\ref{e:msdiag z2}), we may write,
\begin{eqnarray}\label{e:MS1}
	{\cal M}_{S}^2  &\equiv& {\cal O}_\alpha^T  \cdot {\rm diag} (m_h^2,~m_{H_1}^2,~ m_{H_2}^2) \cdot {\cal O}_\alpha\,,
\end{eqnarray}
which facilitates the solution of the remaining six lambdas as follows:
\begin{subequations}
	\label{e:lam1to6 z2 scale}
	\begin{eqnarray}
		\lambda_1  &=& \frac{m_h^2}{2 v^2} \frac{c^2_{\alpha_1} c^2_{\alpha_2} }{c^2_{\beta_1} c^2_{\beta_2}}
		+ \frac{1}{2 v^2 c^2_{\beta_1} c^2_{\beta_2} }
		\left[
		m_{H_1}^2\left(s_{\alpha_1} c_{\alpha_3}- c_{\alpha_1} s_{\alpha_2}  s_{\alpha_3}  \right)^2 -M_1^2\left(s_{\beta_1} c_{\alpha}- c_{\beta_1} s_{\beta_2}  s_{\alpha}\right)^2 \right]\nonumber\\
		&&
		+  \frac{1}{2 v^2 c^2_{\beta_1}c^2_{\beta_2}} \left[m_{H_2}^2\left(c_{\alpha_1}   s_{\alpha_2}  c_{\alpha_3} + s_{\alpha_1} s_{\alpha_3}\right)^2-M_2^2\left(c_{\beta_1}  s_{\beta_2}  c_{\alpha} + s_{\beta_1} s_{\alpha}\right)^2\right]\,,\\
		\lambda_2  &=& \frac{m_h^2}{2 v^2}\frac{s^2_{\alpha_1} c^2_{\alpha_2}  } {s^2_{\beta_1}c^2_{\beta_2}  }
		+\frac{1}{2 v^2 s^2_{\beta_1} c^2_{\beta_2}}\left[m_{H_1}^2\left(c_{\alpha_1} c_{\alpha_3} + s_{\alpha_1}s_{\alpha_2} s_{\alpha_3} \right)^2 -M_1^2 \left(c_{\beta_1} c_{\alpha} + s_{\beta_1}s_{\beta_2} s_{\alpha} \right)^2 \right]\nonumber\\
		&&
		+\frac{1}{2 v^2 s^2_{\beta_1} c^2_{\beta_2}} \left[m_{H_2}^2\left(c_{\alpha_1} s_{\alpha_3} - s_{\alpha_1}s_{\alpha_2} c_{\alpha_3} \right)^2 -M_2^2 \left(c_{\beta_1} s_{\alpha} - s_{\beta_1}s_{\beta_2} c_{\alpha} \right)^2    \right]  \,, \\
		\lambda_3  &=& \frac{m_h^2}{2v^2} \frac{s^2_{\alpha_2}}{s^2_{\beta_2}}
		+\frac{1}{2v^2 s^2_{\beta_2}}\left[m_{H_1}^2 c^2_{\alpha_2} s^2_{\alpha_3}
		+m_{H_2}^2 c^2_{\alpha_2} c^2_{\alpha_3} -M_1^2 c^2_{\beta_2}s^2_{\alpha} -M_2^2 c^2_{\beta_2} c^2_{\alpha}\right]\,,  \\
		\lambda_4  &=& \frac{m_h^2}{v^2}\frac{  s_{2\alpha_1}c^2_{\alpha_2}}{s_{2\beta_1} c^2_{\beta_2}} +
		\frac{1}{4v^2 s_{2\beta_1} c^2_{\beta_2}}\Big[\left(m_{H_1}^2-m_{H_2}^2\right) \left\{(-3 + c_{2\alpha_2})s_{2\alpha_1}c_{2\alpha_3} + 4 c_{2\alpha_1} s_{\alpha_2} s_{2\alpha_3}\right\}\nonumber \\
		&&
		-\left(M_1^2-M_2^2\right) \left\{(-3 + c_{2\beta_2})s_{2\beta_1}c_{2\alpha}+ 4 c_{2\beta_1} s_{\beta_2} s_{2\alpha}\right\}-2\big\{\left(m_{H_1}^2+m_{H_2}^2\right)s_{2\alpha_1} c^2_{\alpha_2}\nonumber \\
		&&
		 - \left(M_1^2+M_2^2\right)s_{2\beta_1} c^2_{\beta_2}\big\} \Big]  
		-\lambda_{7}-2\lambda_{10}  \,, \\
		\lambda_5 &=&\frac{m_h^2}{ v^2 } \frac{c_{\alpha_1} s_{2\alpha_2}}{c_{\beta_1} s_{2\beta_2}}
		+\frac{1}{v^2c_{\beta_1} s_{2\beta_2}}\Big[\left(m_{H_1}^2 -m_{H_2}^2\right) s_{\alpha_1} c_{\alpha_2} s_{2\alpha_3}-\left(M_1^2-M_2^2\right) s_{\beta_1} c_{\beta_2} s_{2\alpha}
		\nonumber \\
		&& -\left(m_{H_1}^2 s^2_{\alpha_3} +m_{H_2}^2 c^2_{\alpha_3} \right) c_{\alpha_1}  s_{2 \alpha_2}
		+ \left(M_1^2 s^2_{\alpha} +M_2^2 c^2_{\alpha} \right)c_{\beta_1} s_{2\beta_2} \Big] - \lambda_8 -2\lambda_{11}   \,,  \\
		\lambda_6  &=&  
		\frac{ m_h^2 }{ v^2}\frac{s_{\alpha_1}s_{2\alpha_2} } {s_{\beta_1} s_{2\beta_2}} 
		+ \frac{1}{v^2 s_{\beta_1} s_{2\beta_2}}
		\Big[
		\left(m_{H_2}^2-m_{H_1}^2\right) c_{\alpha_1} c_{\alpha_2} s_{2 \alpha_3} - \left(M_2^2 -M_1^2\right) c_{\beta_1} c_{\beta_2} s_{2\alpha} \nonumber \\
		&& -\left(m_{H_1}^2 s^2_{\alpha_3} +m_{H_2}^2 c^2_{\alpha_3}\right) s_{\alpha_1} s_{2 \alpha_2} 
		+\left(M_1^2 s^2_{\alpha} + M_2^2 c^2_{\alpha}\right)   s_{\beta_1} s_{2 \beta_2} \Big] -\lambda_{9}-2\lambda_{12}  \,.
	\end{eqnarray} 
\end{subequations}
In these expressions, $\lambda_{7-12}$ can be replaced using Eq.~(\ref{e:lam12 z2 scales}) and~(\ref{e:lam789 z2}). 

Having derived the expressions for all twelve quartic parameters $(\lambda_{1-12})$ in terms of the physical masses and mixings, we can now proceed to determine the scalar self couplings involving the physical scalar fields.
Of particular interest are the interactions between the SM-like Higgs ($h$) and the charged scalar states ($H_{1,2}^\pm$).  
In the alignment limit, these couplings are given as,
%Now that we have obtained the expressions for all twelve quartic couplings ($\lambda s$). we can find the couplings of the physical scalars. In the alignment limit, lightest CP-even Higgs will exactly mimic the SM Higgs at tree level. So the coupling of SM Higgs to the charged scalars can be expressed as 
\begin{subequations}
	\label{e:ghch-alignment}
\begin{eqnarray}
	g_{h H^{+}_{1} H^{-}_{1}} &=& -\frac{1}{v}\left\{2 m^2_{H^{\pm}_{1}}+m^2_{h} - 2 M_1^2 \cos^2(\gamma_2-\alpha) - 2 M_2^2 \sin^2(\gamma_2-\alpha)\right\}\,, \label{e:ghCh1-alignment} \\
		g_{h H^{+}_{2} H^{-}_{2}} &=& -\frac{1}{v}\left \{2 m^2_{H^{\pm}_{2}}+m^2_{h} - 2 M_2^2 \cos^2(\gamma_2-\alpha) - 2 M_1^2 \sin^2(\gamma_2-\alpha)\right\}\,. \label{e:ghCh2-alignment}
\end{eqnarray}
\end{subequations}
%\begin{eqnarray}
%	g_{h H^{+}_{i} H^{-}_{i}} &=& -\frac{-2 m^2_{H^{\pm}_{i}}-m^2_{h} + 2M^2_1 \cos^2 (\gamma_2-\alpha) + 2 M^2_2\sin^2 (\gamma_2-\alpha)}{v}
%\end{eqnarray}

We now consider the impact of the limit of Eq.~(\ref{e:theolimit}) on the scalar potential, especially when applied in conjunction with alignment limit.
When \Eqn{e:theolimit} together with \Eqn{e:alignment} are substituted into the expressions of  Eqs. (\ref{e:lam12 z2 scales}), (\ref{e:lam789 z2}) and (\ref{e:lam1to6 z2 scale}), the expressions for the lambdas get simplified as follows
\begin{subequations}
		\label{e:lambda_theolimit}
\begin{eqnarray}
	\lambda_1=\lambda_2=\lambda_3
	=\frac{\lambda_4}{2}
	=\frac{\lambda_5}{2}
	=\frac{\lambda_6}{2}
	=\frac{m_h^2}{2v^2} \equiv \lambda ~(\rm say) \, , \\
	\lambda_7 =\lambda_8=\lambda_9=\lambda_{10} =\lambda_{11}=\lambda_{12}\;=0\; .
\end{eqnarray}
\end{subequations}
%And accordingly for the coupling of SM higgs to the charged scalars becomes 
%\begin{eqnarray}
%	g_{h H^{+}_{i} H^{-}_{i}} &=& -\frac{m^2_{h}}{v}
%\end{eqnarray}
%Thus, in this limit, quartic couplings can be completely determined by the SM-Higgs mass and the EW VEV which are stable under perturbativity, unitarity and other theoretical constraints. Accordingly scalar potential takes a simplified form with an enhanced symmetry as follows:
Consequently, the scalar potential of \Eqn{e:potential_z2} assumes a much simpler form as given below~\cite{Chakraborti:2021bpy}
\begin{eqnarray}
	V_{Z_2 \times Z_2^\prime} &=& m_{11}^2(\phi_{\bl}^\dagger \phi_{\bl}) +  m_{22}^2(\phi_{\bd}^\dagger 
	\phi_{\bd}) +  m_{33}^2 (\phi_{\bu}^\dagger \phi_{\bu})  - \left(m_{12}^2(\phi_{\bl}^\dagger \phi_{\bd}) +  m_{23}^2(\phi_{\bd}^\dagger 
	\phi_{\bu}) +  m_{13}^2 (\phi_{\bl}^\dagger \phi_{\bu}) + {\rm h.c.} \right)   \nonumber \\
	&&  +\lambda(\phi_{\bl}^\dagger \phi_{\bl} +\phi_{\bd}^\dagger \phi_{\bd}+\phi_{\bu}^\dagger \phi_{\bu})^2 \,.	%	
	\label{e:potential_sp6}
\end{eqnarray}

This potential exhibits an $Sp(6)$ symmetry~\cite{Darvishi:2021txa} in the quartic part which is a much larger symmetry than the $Z_2 \times Z_2^\prime$ symmetry of 
\Eqn{e:potential_z2}. Moreover, the only quartic parameter ($\lambda$) in \Eqn{e:potential_sp6} is determined as in \Eqn{e:lambda_theolimit}. Since 
$m_h = 125 $ GeV and $v = 246$ GeV, this quartic coupling satisfies the limit $0 < \lambda < 8 \pi$. In view of this,
we can conclude that the scalar potential of \Eqn{e:potential_sp6} is automatically consistent with the theoretical constraints arising from unitarity and BFB.
We also note that, in this limit, the couplings of \Eqn{e:ghch-alignment} take a particularly simpler form as
\begin{eqnarray}
	g_{h H^{+}_{1} H^{-}_{1}} = g_{h H^{+}_{2} H^{-}_{2}} &=& -\frac{m^2_{h}}{v} \,.
	\label{e:ghch-theolimit}
\end{eqnarray}

%% file: v2.bbl
\providecommand{\href}[2]{#2}\begingroup\raggedright\begin{thebibliography}{10}

\bibitem{Branco:2011iw}
G.~C. Branco, P.~M. Ferreira, L.~Lavoura, M.~N. Rebelo, M.~Sher and J.~P.
  Silva, \emph{{Theory and phenomenology of two-Higgs-doublet models}},
  \href{http://dx.doi.org/10.1016/j.physrep.2012.02.002}{\emph{Phys. Rept.}
  {\bf 516} (2012) 1--102}, [\href{https://arxiv.org/abs/1106.0034}{{\tt
  1106.0034}}].

\bibitem{Bhattacharyya:2015nca}
G.~Bhattacharyya and D.~Das, \emph{{Scalar sector of two-Higgs-doublet models:
  A minireview}},
  \href{http://dx.doi.org/10.1007/s12043-016-1252-4}{\emph{Pramana} {\bf 87}
  (2016) 40}, [\href{https://arxiv.org/abs/1507.06424}{{\tt 1507.06424}}].

\bibitem{Cree:2011uy}
G.~Cree and H.~E. Logan, \emph{{Yukawa alignment from natural flavor
  conservation}},
  \href{http://dx.doi.org/10.1103/PhysRevD.84.055021}{\emph{Phys. Rev. D} {\bf
  84} (2011) 055021}, [\href{https://arxiv.org/abs/1106.4039}{{\tt
  1106.4039}}].

\bibitem{Das:2019yad}
D.~Das and I.~Saha, \emph{{Alignment limit in three Higgs-doublet models}},
  \href{http://dx.doi.org/10.1103/PhysRevD.100.035021}{\emph{Phys. Rev. D} {\bf
  100} (2019) 035021}, [\href{https://arxiv.org/abs/1904.03970}{{\tt
  1904.03970}}].

\bibitem{Chakraborti:2021bpy}
M.~Chakraborti, D.~Das, M.~Levy, S.~Mukherjee and I.~Saha, \emph{{Prospects for
  light charged scalars in a three-Higgs-doublet model with Z3 symmetry}},
  \href{http://dx.doi.org/10.1103/PhysRevD.104.075033}{\emph{Phys. Rev. D} {\bf
  104} (2021) 075033}, [\href{https://arxiv.org/abs/2104.08146}{{\tt
  2104.08146}}].

\bibitem{Boto:2021qgu}
R.~Boto, J.~C. Rom{\~a}o and J.~P. Silva, \emph{{Current bounds on the type-Z
  Z3 three-Higgs-doublet model}},
  \href{http://dx.doi.org/10.1103/PhysRevD.104.095006}{\emph{Phys. Rev. D} {\bf
  104} (2021) 095006}, [\href{https://arxiv.org/abs/2106.11977}{{\tt
  2106.11977}}].

\bibitem{Porto:2007ed}
R.~A. Porto and A.~Zee, \emph{{The Private Higgs}},
  \href{http://dx.doi.org/10.1016/j.physletb.2008.08.001}{\emph{Phys. Lett. B}
  {\bf 666} (2008) 491--495}, [\href{https://arxiv.org/abs/0712.0448}{{\tt
  0712.0448}}].

\bibitem{Porto:2008hb}
R.~A. Porto and A.~Zee, \emph{{Neutrino Mixing and the Private Higgs}},
  \href{http://dx.doi.org/10.1103/PhysRevD.79.013003}{\emph{Phys. Rev. D} {\bf
  79} (2009) 013003}, [\href{https://arxiv.org/abs/0807.0612}{{\tt
  0807.0612}}].

\bibitem{Glashow:1976nt}
S.~L. Glashow and S.~Weinberg, \emph{{Natural Conservation Laws for Neutral
  Currents}}, \href{http://dx.doi.org/10.1103/PhysRevD.15.1958}{\emph{Phys.
  Rev. D} {\bf 15} (1977) 1958}.

\bibitem{ATLAS:2020zms}
{\scshape ATLAS} collaboration, G.~Aad et~al., \emph{{Search for heavy Higgs
  bosons decaying into two tau leptons with the ATLAS detector using $pp$
  collisions at $\sqrt{s}=13$ TeV}},
  \href{http://dx.doi.org/10.1103/PhysRevLett.125.051801}{\emph{Phys. Rev.
  Lett.} {\bf 125} (2020) 051801},
  [\href{https://arxiv.org/abs/2002.12223}{{\tt 2002.12223}}].

\bibitem{Ivanov:2011ae}
I.~P. Ivanov, V.~Keus and E.~Vdovin, \emph{{Abelian symmetries in
  multi-Higgs-doublet models}},
  \href{http://dx.doi.org/10.1088/1751-8113/45/21/215201}{\emph{J. Phys. A}
  {\bf 45} (2012) 215201}, [\href{https://arxiv.org/abs/1112.1660}{{\tt
  1112.1660}}].

\bibitem{Keus:2013hya}
V.~Keus, S.~F. King and S.~Moretti, \emph{{Three-Higgs-doublet models:
  symmetries, potentials and Higgs boson masses}},
  \href{http://dx.doi.org/10.1007/JHEP01(2014)052}{\emph{JHEP} {\bf 01} (2014)
  052}, [\href{https://arxiv.org/abs/1310.8253}{{\tt 1310.8253}}].

\bibitem{Ivanov:2014doa}
I.~P. Ivanov and C.~C. Nishi, \emph{{Symmetry breaking patterns in 3HDM}},
  \href{http://dx.doi.org/10.1007/JHEP01(2015)021}{\emph{JHEP} {\bf 01} (2015)
  021}, [\href{https://arxiv.org/abs/1410.6139}{{\tt 1410.6139}}].

\bibitem{Das:2022gbm}
D.~Das, M.~Levy, P.~B. Pal, A.~M. Prasad, I.~Saha and A.~Srivastava,
  \emph{{Democratic three-Higgs-doublet models: The custodial limit and
  wrong-sign Yukawa coupling}},
  \href{http://dx.doi.org/10.1103/PhysRevD.107.055035}{\emph{Phys. Rev. D} {\bf
  107} (2023) 055035}, [\href{https://arxiv.org/abs/2301.00231}{{\tt
  2301.00231}}].

\bibitem{Boto:2023nyi}
R.~Boto, D.~Das, L.~Lourenco, J.~C. Romao and J.~P. Silva,
  \emph{{Fingerprinting the type-Z three-Higgs-doublet models}},
  \href{http://dx.doi.org/10.1103/PhysRevD.108.015020}{\emph{Phys. Rev. D} {\bf
  108} (2023) 015020}, [\href{https://arxiv.org/abs/2304.13494}{{\tt
  2304.13494}}].

\bibitem{Boto:2024jgj}
R.~Boto, L.~Lourenco, J.~C. Rom{\~a}o and J.~P. Silva, \emph{{Large
  pseudoscalar Yukawa couplings in the complex 3HDM}},
  \href{http://dx.doi.org/10.1007/JHEP11(2024)106}{\emph{JHEP} {\bf 11} (2024)
  106}, [\href{https://arxiv.org/abs/2407.19856}{{\tt 2407.19856}}].

\bibitem{Darvishi:2021txa}
N.~Darvishi, M.~R. Masouminia and A.~Pilaftsis, \emph{{Maximally symmetric
  three-Higgs-doublet model}},
  \href{http://dx.doi.org/10.1103/PhysRevD.104.115017}{\emph{Phys. Rev. D} {\bf
  104} (2021) 115017}, [\href{https://arxiv.org/abs/2106.03159}{{\tt
  2106.03159}}].

\bibitem{Bento:2017eti}
M.~P. Bento, H.~E. Haber, J.~C. Rom{\~a}o and J.~P. Silva, \emph{{Multi-Higgs
  doublet models: physical parametrization, sum rules and unitarity bounds}},
  \href{http://dx.doi.org/10.1007/JHEP11(2017)095}{\emph{JHEP} {\bf 11} (2017)
  095}, [\href{https://arxiv.org/abs/1708.09408}{{\tt 1708.09408}}].

\bibitem{Bento:2022vsb}
M.~P. Bento, J.~C. Rom{\~a}o and J.~P. Silva, \emph{{Unitarity bounds for all
  symmetry-constrained 3HDMs}},
  \href{http://dx.doi.org/10.1007/JHEP08(2022)273}{\emph{JHEP} {\bf 08} (2022)
  273}, [\href{https://arxiv.org/abs/2204.13130}{{\tt 2204.13130}}].

\bibitem{Boto:2022uwv}
R.~Boto, J.~C. Rom{\~a}o and J.~P. Silva, \emph{{Bounded from below conditions
  on a class of symmetry constrained 3HDM}},
  \href{http://dx.doi.org/10.1103/PhysRevD.106.115010}{\emph{Phys. Rev. D} {\bf
  106} (2022) 115010}, [\href{https://arxiv.org/abs/2208.01068}{{\tt
  2208.01068}}].

\bibitem{Jurciukonis:2026dmc}
D.~Jur{\v{c}}iukonis, L.~Lavoura and A.~Milagre, \emph{{Assessing boundedness
  from below in the $\mathbb{Z}_2 \times \mathbb{Z}_2$-symmetric
  three-Higgs-doublet model: algorithm and machine learning}},
  \href{http://dx.doi.org/10.21468/SciPostPhysCore.9.3.041}{\emph{SciPost Phys.
  Core} {\bf 9} (2026) 041}, [\href{https://arxiv.org/abs/2603.23590}{{\tt
  2603.23590}}].

\bibitem{Alloul:2013bka}
A.~Alloul, N.~D. Christensen, C.~Degrande, C.~Duhr and B.~Fuks,
  \emph{{FeynRules 2.0 - A complete toolbox for tree-level phenomenology}},
  \href{http://dx.doi.org/10.1016/j.cpc.2014.04.012}{\emph{Comput. Phys.
  Commun.} {\bf 185} (2014) 2250--2300},
  [\href{https://arxiv.org/abs/1310.1921}{{\tt 1310.1921}}].

\bibitem{Degrande:2011ua}
C.~Degrande, C.~Duhr, B.~Fuks, D.~Grellscheid, O.~Mattelaer and T.~Reiter,
  \emph{{UFO - The Universal FeynRules Output}},
  \href{http://dx.doi.org/10.1016/j.cpc.2012.01.022}{\emph{Comput. Phys.
  Commun.} {\bf 183} (2012) 1201--1214},
  [\href{https://arxiv.org/abs/1108.2040}{{\tt 1108.2040}}].

\bibitem{Alwall:2014hca}
J.~Alwall, R.~Frederix, S.~Frixione, V.~Hirschi, F.~Maltoni, O.~Mattelaer
  et~al., \emph{{The automated computation of tree-level and next-to-leading
  order differential cross sections, and their matching to parton shower
  simulations}}, \href{http://dx.doi.org/10.1007/JHEP07(2014)079}{\emph{JHEP}
  {\bf 07} (2014) 079}, [\href{https://arxiv.org/abs/1405.0301}{{\tt
  1405.0301}}].

\bibitem{ParticleDataGroup:2026aaa}
{\scshape Particle Data Group} collaboration, F.~Takahashi et~al.,
  \emph{{Review of Particle Physics}},
  \href{http://dx.doi.org/10.1142/S0217751X26300115}{\emph{Int. J. Mod. Phys.
  A} {\bf 41} (2026) 2630011}.

\bibitem{ATLAS:2026pdi}
{\scshape ATLAS} collaboration, G.~Aad et~al., \emph{{Combination of Higgs
  boson measurements at $\sqrt{s} =$ 13 TeV and their interpretations by the
  ATLAS experiment}},  \href{https://arxiv.org/abs/2608.07332}{{\tt
  2608.07332}}.

\bibitem{Kanemura:2014dea}
S.~Kanemura, H.~Yokoya and Y.-J. Zheng, \emph{{Complementarity in direct
  searches for additional Higgs bosons at the LHC and the International Linear
  Collider}},
  \href{http://dx.doi.org/10.1016/j.nuclphysb.2014.07.00}{\emph{Nucl. Phys. B}
  {\bf 886} (2014) 524--553}, [\href{https://arxiv.org/abs/1404.5835}{{\tt
  1404.5835}}].

\bibitem{Aiko:2020ksl}
M.~Aiko, S.~Kanemura, M.~Kikuchi, K.~Mawatari, K.~Sakurai and K.~Yagyu,
  \emph{{Probing extended Higgs sectors by the synergy between direct searches
  at the LHC and precision tests at future lepton colliders}},
  \href{http://dx.doi.org/10.1016/j.nuclphysb.2021.115375}{\emph{Nucl. Phys. B}
  {\bf 966} (2021) 115375}, [\href{https://arxiv.org/abs/2010.15057}{{\tt
  2010.15057}}].

\bibitem{Cepeda:2019klc}
M.~Cepeda et~al., \emph{{Report from Working Group 2}: {Higgs Physics at the
  HL-LHC and HE-LHC}},
  \href{http://dx.doi.org/10.23731/CYRM-2019-007.221}{\emph{CERN Yellow Rep.
  Monogr.} {\bf 7} (2019) 221--584},
  [\href{https://arxiv.org/abs/1902.00134}{{\tt 1902.00134}}].

\bibitem{ATLAS:2024hya}
{\scshape ATLAS} collaboration, G.~Aad et~al., \emph{{Search for charged Higgs
  bosons produced in top-quark decays or in association with top quarks and
  decaying via
  H{\ensuremath{\pm}}{\textrightarrow}{\ensuremath{\tau}}{\ensuremath{\pm}}{\ensuremath{\nu}}{\ensuremath{\tau}}
  in 13~TeV pp collisions with the ATLAS detector}},
  \href{http://dx.doi.org/10.1103/PhysRevD.111.072006}{\emph{Phys. Rev. D} {\bf
  111} (2025) 072006}, [\href{https://arxiv.org/abs/2412.17584}{{\tt
  2412.17584}}].

\bibitem{CMS:2025frm}
{\scshape CMS} collaboration, V.~Chekhovsky et~al., \emph{{Search for bosons of
  an extended Higgs sector in b quark final states in proton-proton collisions
  at $ \sqrt{\textrm{s}} $ = 13 TeV}},
  \href{http://dx.doi.org/10.1007/JHEP06(2025)144}{\emph{JHEP} {\bf 06} (2025)
  144}, [\href{https://arxiv.org/abs/2502.06568}{{\tt 2502.06568}}].

\bibitem{ATLAS:2021upq}
{\scshape ATLAS} collaboration, G.~Aad et~al., \emph{{Search for charged Higgs
  bosons decaying into a top quark and a bottom quark at $ \sqrt{\mathrm{s}} $
  = 13 TeV with the ATLAS detector}},
  \href{http://dx.doi.org/10.1007/JHEP06(2021)145}{\emph{JHEP} {\bf 06} (2021)
  145}, [\href{https://arxiv.org/abs/2102.10076}{{\tt 2102.10076}}].

\bibitem{Weinberg:1976hu}
S.~Weinberg, \emph{{Gauge Theory of CP Violation}},
  \href{http://dx.doi.org/10.1103/PhysRevLett.37.657}{\emph{Phys. Rev. Lett.}
  {\bf 37} (1976) 657}.

\bibitem{Plantey:2022jdg}
R.~Plantey, O.~M. Ogreid, P.~Osland, M.~N. Rebelo and M.~A. Solberg,
  \emph{{Weinberg{\textquoteright}s 3HDM potential with spontaneous CP
  violation}}, \href{http://dx.doi.org/10.1103/PhysRevD.108.075029}{\emph{Phys.
  Rev. D} {\bf 108} (2023) 075029},
  [\href{https://arxiv.org/abs/2208.13594}{{\tt 2208.13594}}].

\bibitem{Ogreid:2024qfw}
O.~M. Ogreid, P.~Osland and M.~N. Rebelo, \emph{{CP-violation in the Weinberg
  3HDM potential}},
  \href{http://dx.doi.org/10.1007/JHEP09(2025)111}{\emph{JHEP} {\bf 09} (2025)
  111}, [\href{https://arxiv.org/abs/2411.05480}{{\tt 2411.05480}}].

\bibitem{Bhattacharyya:2014oka}
G.~Bhattacharyya and D.~Das, \emph{{Nondecoupling of charged scalars in Higgs
  decay to two photons and symmetries of the scalar potential}},
  \href{http://dx.doi.org/10.1103/PhysRevD.91.015005}{\emph{Phys. Rev. D} {\bf
  91} (2015) 015005}, [\href{https://arxiv.org/abs/1408.6133}{{\tt
  1408.6133}}].

\bibitem{Faro:2020qyp}
F.~Faro, J.~C. Romao and J.~P. Silva, \emph{{Nondecoupling in Multi-Higgs
  doublet models}},
  \href{http://dx.doi.org/10.1140/epjc/s10052-020-8217-y}{\emph{Eur. Phys. J.
  C} {\bf 80} (2020) 635}, [\href{https://arxiv.org/abs/2002.10518}{{\tt
  2002.10518}}].

\bibitem{Das:2026ldf}
D.~Das, M.~Levy, S.~Pandey, I.~Saha and A.~Sarkar, \emph{{Soft Symmetry
  Breaking as a Nonstandard Source of Mass: Phenomenological Insights from the
  Two-Higgs-Doublet Model}},  \href{https://arxiv.org/abs/2603.23280}{{\tt
  2603.23280}}.

\end{thebibliography}\endgroup
